\documentclass[a4paper,10pt,3p,twocolumn]{elsarticle}

\usepackage{graphicx}
\usepackage{hyperref}
\usepackage{siunitx}
\usepackage[version=4]{mhchem}
\usepackage{caption}
\usepackage{subcaption}
\usepackage{amsmath}
\journal{Astroparticle Physics}
\begin{document}

\begin{frontmatter}

\title{\boldmath Production and propagation of ultra-high energy photons using CRPropa 3}

\author[a,b]{C.~Heiter}
\author[a,c]{D.~Kuempel}
\author[a]{D.~Walz}
\author[a]{M.~Erdmann}
\ead{erdmann@physik.rwth-aachen.de}

\address[a]{RWTH Aachen University, III. Physikalisches Institut A, Otto-Blumenthal-Str., 52056 Aachen, Germany}
\address[b]{Max Planck Institute for Radio Astronomy, Auf dem H\"ugel 69, 53121 Bonn, Germany}
\address[c]{Bergische Universit\"at Wuppertal, Department of Physics, Gau\ss str. 20, 42097 Wuppertal, Germany}

\begin{abstract}
In order to interpret cosmic ray observations, detailed modeling of propagation effects invoking all important 
messengers is necessary.
We introduce a new photon production and propagation code as an inherent part of the CRPropa~3 software framework.
By implementing additional photon production channels, which are important for energies below $\sim \SI{e18}{eV}$, 
this code can be used for multi-messenger studies connecting the TeV and sub EeV energy regime and for interpreting 
models of ultra-high energy cosmic ray sources.
We discuss the importance of the individual production channels and propagation effects and present example applications.
\end{abstract}


\end{frontmatter}

\section{Introduction}

Photons are unique messengers for exploring the universe.
The currently maximum energy of an identified photon is in the TeV energy regime detected by ground-based Cherenkov telescopes.
At higher energies only charged cosmic rays and neutrinos have been observed.
The origin of ultra-high energy (UHE) cosmic rays with energies above $\sim \SI{1}{EeV}$ is a long-standing mystery and large-scale observatories have been built (e.g.\ \cite{ThePierreAuger:2015rma,TARefAbuZayyad201287,Tokuno201254,IceCubeRefAchterberg2006155}) to locate the sources and to study the acceleration mechanism that provides for producing particles with such high energies.
There is a correlation between acceleration sites of UHE cosmic rays and the emission of high-energy photons and neutrinos.
The observation and interpretation of these particles in different energy regimes in a so-called multi-messenger approach allows for maximizing the scientific information and will increase the chances of identifying the sources of UHE cosmic rays.

Above the GeV energy range, photons cannot conceivably be generated by thermal emission from hot celestial objects.
Instead, UHE photons probe the non-thermal Universe where other mechanisms allow the concentration of large amounts of energy into a single particle.
Typically, the production is associated with the decay of a neutral pion previously produced by a `primary' process.
In conventional astrophysical models this process is the interaction of UHE cosmic rays with low-energy background photons in a GZK-type process \cite{GreisenPhysRevLett,Zatsepin:1966jv} of resonant photo-pion production (with a final photon energy typically a factor of 10 below the energy of the primary nucleon \cite{Risse:2007sd}) or proton-proton interactions.
Depending on the energy of interest, other production mechanisms may become important, such as photons from photodisintegration, elastic scattering, pair-production processes, radiative decay or synchrotron emission (cf.\ Sec.\ \ref{sec:production}).

Other predictions of UHE photons arise from non-acceleration models where the primary process is given by the decay (or annihilation) of primordial relics such as topological defects \cite{HILL1983469,Hindmarsh0034}, super-heavy dark matter \cite{PhysRevLett.79.4302,BIRKEL1998297,Kuzmin1998,Blasi200257,PhysRevD.94.063535}, $Z$-burst scenarios \cite{WeilerPhysRevLett,WEILER1999303,Fargon0004} or other top-down models \cite{PhysRevD.74.115003,FodorPhysRevLett,Sarkar2002495,Barbot20035,AloisioPhysRev}.
However, several experiments have searched for a flux of these photons and have set limits that severely constrain the parameter space of non-acceleration models \cite{BleveICRC,Aab:2016agp,Rubtsov:2015trh}.
With the advent of large-scale observatories such as the Pierre Auger Observatory \cite{ThePierreAuger:2015rma} or Telescope Array \cite{TARefAbuZayyad201287,Tokuno201254}, a great wealth of data of unprecedented quality and quantity is now being accumulated and flux limits already constrain optimistic predictions of GZK-type processes \cite{Aab:2016agp}.
The predictions are strongly dependent on the chemical composition at the highest energies.
At energies around the `ankle' of the energy spectrum at about \SI{5}{EeV}~\cite{Spectrum, ICRC2013_S}, several experiments, including the Pierre Auger Observatory, HiRes, and Telescope Array, have all found their measurements to be consistent with a large fraction of light elements among the cosmic rays \cite{xsection, Xmax, HiRes, TA}.
At higher energies, composition measurements indicate an increasing fraction of heavier nuclei reducing the predicted flux of UHE photons \cite{PhysRevD.90.122006,Aab2016288}.

By comparing the information provided by multiple messengers in combination with accurate Monte Carlo predictions, astrophysical scenarios can be tested.
In order to interpret measurements from large-scale observatories it is therefore essential to have a general and versatile simulation tool for the propagation of cosmic rays.
In this context, CRPropa~3~\cite{CRPropa3_Paper} was developed to enable the production and propagation of cosmic rays and their secondaries in a configurable time-dependent environment in three spatial dimensions.

In this paper we present two major new implementations, incorporated in the CRPropa~3.1 release, regarding the production and propagation of high-energy photons.
First, we present and discuss several additional production channels for photons by UHE cosmic-ray nuclei, which were previously not considered.
Second, we present an implementation of the propagation of electromagnetic (EM) particles as an integral part of the CRPropa framework.
This implementation enables the propagation in structured magnetic and background photon fields.
The new code can be used to study multi-messenger aspects of UHE cosmic rays, e.g., computing the expected UHE photon fluxes from various astrophysical scenarios or connecting TeV photon observations to UHE cosmic-ray physics.

This paper is structured as follows.
We first give a brief introduction to the CRPropa framework in section \ref{sec:crpropa}. 
In section \ref{sec:production} we describe the relevant EM production processes of UHE nuclei, focusing on the newly implemented channels.
Then, in section \ref{sec:propagation} we describe the implementation of the EM propagation in CRPropa.
In section \ref{sec:results} we assess the importance of the individual production channels at the point of creation (section \ref{sec:results-budget}), as well as in the observable photon fluxes after propagation (section \ref{sec:results-propagation}).
Finally, in section \ref{sec:results-horizon} we apply CRPropa to calculate the expected photon horizons for different scenarios before providing a summary in section \ref{sec:summary}.
\section{The CRPropa framework}
\label{sec:crpropa}

Within the CRPropa framework, ultra-relativistic particles can be propagated through galactic and extragalactic space.
CRPropa is a flexible, public code built to simulate the energy spectrum, mass composition and arrival direction distribution of cosmic rays for various source arrangements and injection characteristics while accounting for interactions with background photons, deflections in the intervening magnetic fields and cosmological effects.
It is written in \texttt{C++} but can be steered with Python.
The modular design enables the user to customize the simulation by combining existing simulation modules with own modules written in Python or \texttt{C++}.

Particles are injected with initial properties such as particle type, position, momentum and time, and are individually propagated in discrete steps of variable size.
Within each step the probability for an interaction is calculated given the interaction length $\lambda$ of a specific process.
The propagation is stopped once a user-defined break condition is met, e.g.\ when a particle reaches the observer or drops below a defined energy threshold.

The particle type is specified via the Particle Data Group numbering scheme \cite{PDBook} which comprises all elementary particles, hadrons and nuclei, including isomeric states, as well as a number of hypothetical particles.
General simulation aspects in CRPropa such as adiabatic energy losses or deflection of charged particles work independently of the particle type.
Hence, extending the simulation to new particle types requires only the implementation of the corresponding interaction processes.
The available interaction processes for cosmic-ray nuclei as well as for photons and electrons and positrons (electrons hereafter) on various target fields are listed in Table \ref{tab:Interactions} together with possible secondary particles.
The list includes the newly implemented photon production channels by nuclei, described in section \ref{sec:production}, and the interaction processes of EM particles, described in section \ref{sec:propagation}.

\begin{table*}[htb]
\centering
	\begin{tabular}{ c | c l c }
	\hline \hline
	Initial state & Target field & Process & Secondaries \\
	\hline
	Nuclei & CBR & Pair production (Bethe-Heitler) & $e^\pm$ \\
	Nuclei & CBR & Photo-pion production & $p,n,\nu,e^\pm,\gamma$ \\
	Nuclei & CBR & Photodisintegration & $p,n,d,t, \mathrm{^3He},\alpha,\gamma$* \\
	Nuclei & CBR & Elastic scattering* & $\gamma$ \\
	Nuclei & --  & Nuclear decay & $p,n,\nu,e^\pm,\gamma$* \\
	\hline
	Photons   & CBR & Pair production* (Breit-Wheeler) & $e^\pm$ \\
	Photons   & CBR & Double pair production* & $e^\pm$ \\
	Electrons & CBR & Triplet pair production* & $e^\pm$ \\
	Electrons & CBR & Inverse Compton scattering* & $\gamma$ \\
	Electrons & B-field & Synchrotron radiation* & $\gamma$ \\
	\hline \hline
	\end{tabular}
\caption{
	Implemented processes for cosmic-ray nuclei (including protons and neutrons), photons and electrons interacting with cosmic background radiation (CBR) or with cosmic magnetic fields.
	Secondary particles from these processes are indicated.
	Processes and secondary particles newly implemented in CRPropa 3.1 are denoted by a *.
	Note that while synchrotron radiation applies to all charged particles it is typically not relevant for nuclei.
}
\label{tab:Interactions}
\end{table*}

Low-energy cosmic background radiation fields such as the cosmic microwave background (CMB) form the dominant interaction target for cosmic rays propagating through intergalactic space.
While the spectral shape of the CMB is well known, several models exist for the infrared and optical background (IRB).
The IRB models implemented in CRPropa are Kneiske 2004~\cite{Kneiske2004}, Stecker 2005~\cite{Stecker2005}, Franceschini 2008~\cite{Franceschini2008}, Finke 2010~\cite{Finke2010}, Dominguez 2011~\cite{Dominguez2011}, Gilmore 2012~\cite{Gilmore2012} and, as new additions, the upper and lower bounds determined by Stecker 2016~\cite{Stecker2016}.
Electromagnetic particles at UHE energies can also significantly interact with the cosmic radio background for which the Protheroe~\cite{Protheroe1996} model is implemented.

The interaction rate or inverse interaction length of a cosmic ray of energy $E$, mass $m$ and velocity $\beta \approx 1$ interacting with an isotropic radiation field is given by
\begin{align}
	\lambda^{-1}(E) = \frac{1}{8\beta E^2}  \hspace{-0.1cm}\int_{0}^\infty \hspace{-0.1cm} \frac{n(\epsilon)}{\epsilon^2}
                          \int_{s_\mathrm{th}}^{s_\mathrm{max}}\hspace{-0.1cm} \sigma(s) (s - m^2 c^4) ds \, d\epsilon
	\label{eq:rate}
\end{align}
where $n(\epsilon) = dn/d\epsilon$ is the differential photon number density per energy interval $d\epsilon$ and $\sigma(s)$ is the cross-section of the considered process as a function of the squared center of momentum energy $s$.
A minimum $s_\mathrm{th}$ is required for the production of secondary particles.
The maximum value $s_\mathrm{max} = 4 E \epsilon + m^2$ corresponds to a head-on collision between the cosmic ray and a background photon of energy $\epsilon$.

Several interaction processes described in sections \ref{sec:production} and \ref{sec:propagation} call for sampling the energy of the interacting background photon.
This sampling is done using the differential interaction rate as a probability density function (PDF),
\begin{align}
	\mathrm{PDF}(\epsilon; E) \propto \frac{d \lambda^{-1}}{d \epsilon}(E)~.
	\label{eq:rate-differential}
\end{align}

Since calculating equations (\ref{eq:rate}), (\ref{eq:rate-differential}) is computationally expensive, all (differential) interaction rates are tabulated beforehand and interpolated in CRPropa during runtime.
Here, the interaction rates are calculated separately for each process and each relevant photon background model to enable a fine grained control of the simulation setup.

The redshift dependence of a background field is taken into account through a global scaling factor $s(z)$.
This allows for computing the interaction rates at any redshift from the tabulated values at $z=0$ using the relation
\begin{align*}
	\lambda^{-1}\left(E,z\right) = (1 + z)^2 \, s(z) \, \lambda^{-1}\left(E (1 + z\right), z=0)~.
	\label{eq:rate-redshift}
\end{align*}
This global scaling is exact in the case of the CMB, whereas for the IRB and radio background, whose spectral shapes are redshift-dependent, the error introduced depends on the average propagation distance and is typically small, see \cite{CRPropa3_Paper} for details.
\section{Production of EM particles by UHE nuclei}
\label{sec:production}

In the following we concentrate on important production processes for electromagnetic particles and discuss their implementation in the CRPropa framework.
This includes the existing implementations for photo-pion and pair production, as well as the newly implemented photon production channels via photodisintegration, elastic scattering and radiative decay.
While the additional channels are negligible for cosmic-ray nuclei, they can give rise to a relevant contribution of UHE photons, cf. section \ref{sec:results-budget}.

\subsection{Photo-pion production}
\label{sec:production-photopion}

Pion production for a head-on collision of a nucleon $N$ with a background photon $\gamma_\mathrm{bg}$ can be described by $N + \gamma_\mathrm{bg} \rightarrow N + \pi$, with a high threshold energy of
\begin{align*}
	E_\mathrm{thres}^{N,\pi} &= \frac{m_\pi c^4(m_N + m_\pi /2)}{2\epsilon} \\
                                 &\approx \num{6.8e19}~\left( \frac{\epsilon}{\SI{}{meV}}\right)^{-1}\,\SI{}{eV}~,
\end{align*}
where $m_{\pi}$ and $m_N$ are the masses of the pion and the nucleon and $\epsilon \sim \SI{e-3}{eV}$ represents a typical energy of a CMB photon.
The inelasticity $\eta$ is in the range of $0.2 - 0.5$.
This process is the main production channel of UHE photons and neutrinos by hadronic cosmic rays.
In CRPropa, photo-hadronic processes of protons and neutrons are modeled by the event generator SOPHIA~\cite{SOPHIA} which considers all stable particles that are produced in the interaction.
For nuclei an approximation using a superposition model is used.
The interaction rate of a nucleus is given by
\begin{align*}
	\lambda^{-1}_{A,Z} (E) = 0.85 \cdot \left( Z^\zeta \lambda^{-1}_{p} (E/A) + N^{\zeta} \lambda^{-1}_n (E/A) \right)~,
\end{align*}
where $\lambda^{-1}_p$ and $\lambda^{-1}_n$ are the interaction rates for protons and neutrons and $\zeta = 2/3$ for $A \leq 8$ and $\zeta=1$ for heavier nuclei, cf.\ \cite{CRPropa2}.

\subsection{Pair production (Bethe-Heitler)}
\label{sec:production-pair}

Electron pair production by a nucleus $X$ with mass number $A$ and atomic number $Z$ on a photon $^A_ZX + \gamma_\mathrm{bg} \longrightarrow~ ^A_ZX + e^+ + e^-$ is an important energy loss process for cosmic-ray nuclei.
The threshold energy is two orders of magnitude smaller compared to pion production,
\begin{align*}
	E_\mathrm{thres}^{e^\pm} &= \frac{m_e c^4(m_X + m_e)}{\epsilon} \\
                                 &\approx \num{4.8e17}~A~\left( \frac{\epsilon}{\SI{}{meV}} \right)^{-1}\,\SI{}{eV}~.
\end{align*}
The produced electrons can in turn interact via inverse Compton scattering to produce high-energy secondary photons.
Pair production thus acts as an indirect photon production channel and is especially important when calculating secondary photons below PeV energies.

Electron pair production exhibits the largest cross-section among the photo-hadronic interactions, but has a relatively small inelasticity of about $\eta \sim \num{e-3}$.
As a consequence it is treated as a continuous energy loss process (cf.\ \cite{CRPropa2}).
Given a propagation step size $\Delta s$, the cosmic-ray energy is reduced by $\Delta E = dE / dx \cdot \Delta s$ where the parametrization of the energy loss $dE / dx$ is adopted from \cite{1992PairLoss}.
A number of $e^\pm$-pairs is then produced until their total energy exceeds $\Delta E$, with the last pair being randomly accepted to ensure energy conservation on average.
Here, the energy distribution of the produced $e^\pm$-pairs follows the description given in \cite{PhysRevD.78.034013}.

\subsection{Photodisintegration}
\label{sec:production-disintegration}

Photodisintegration (PD) is the dominant energy loss mechanism for UHE nuclei.
In this process a nucleus interacts inelastically with a cosmic background photon which leads to partial fragmentation of the nucleus.
At photon energies $\epsilon$ of up to \SI{30}{MeV} in the nucleus rest frame, the giant dipole resonance dominates, whereas from 30 to \SI{150}{MeV} the interaction with `quasi-deuterons' within the nucleus becomes important \cite{1976ApJ...205..638P,KHAN2005191}.
It should be noted that, due to the sparsity and quality of the available PD measurements, there are large uncertainties in modeling this process which is especially important for describing nuclei propagation at ultra-high energies \cite{AlvesCRPropaSimProp}.
In CRPropa we use the nuclear reaction code TALYS 1.8 \cite{TALYS,TALYS18} with the settings described in \cite{CRPropa3_Paper} to compute the total cross-section and all relevant branching ratios for the various combinations of nuclear fragments.
The residual nucleus after the PD can be in an excited state, leading to the emission of one or more high-energy photons.
In the following we describe the implementation of this photon emission as new photon production channel in CRPropa 3.1.

As for the PD itself, TALYS 1.8 is used to compute the photon emission branching ratios for all 178 nuclides with $7 \leq A \leq 56$, $Z \leq 26$ and lifetimes $\tau > \SI{2}{s}$.
For the remaining nuclides with $A < \num{7}$, no photon emission data are currently available.
The branching ratios are defined as the quotient of the PD cross-section from mother to excited daughter nucleus with a subsequent emission of a photon of specific energy, and the total cross-section for PD from mother to daughter nucleus.
The branching ratios are tabulated for background photon energies in the nucleus rest frame from 0.2 to \SI{200}{MeV} with 100 log-spaced points per decade.
There are $\num{40325}$ photon emission channels corresponding to the considered PD channels, of which only the $\sim 9500$ most relevant emission channels are selected via a thinning procedure.
Finally, the branching ratios as a function of the nucleus' Lorentz factor are computed by folding with the spectrum of the respective photon background model, cf. equation (\ref{eq:rate}).
The resulting tabulation range covers $\gamma = \num{e6} - \num{e14}$ with 25 log-spaced points per decade.
In case of a PD during a CRPropa simulation, the resulting daughter nucleus is first determined.
Then, a random emission is performed for each photon energy based on the tabulated branching ratios for this daughter nucleus.

\begin{figure}[t]
\begin{center}
\includegraphics[width=0.48\textwidth]{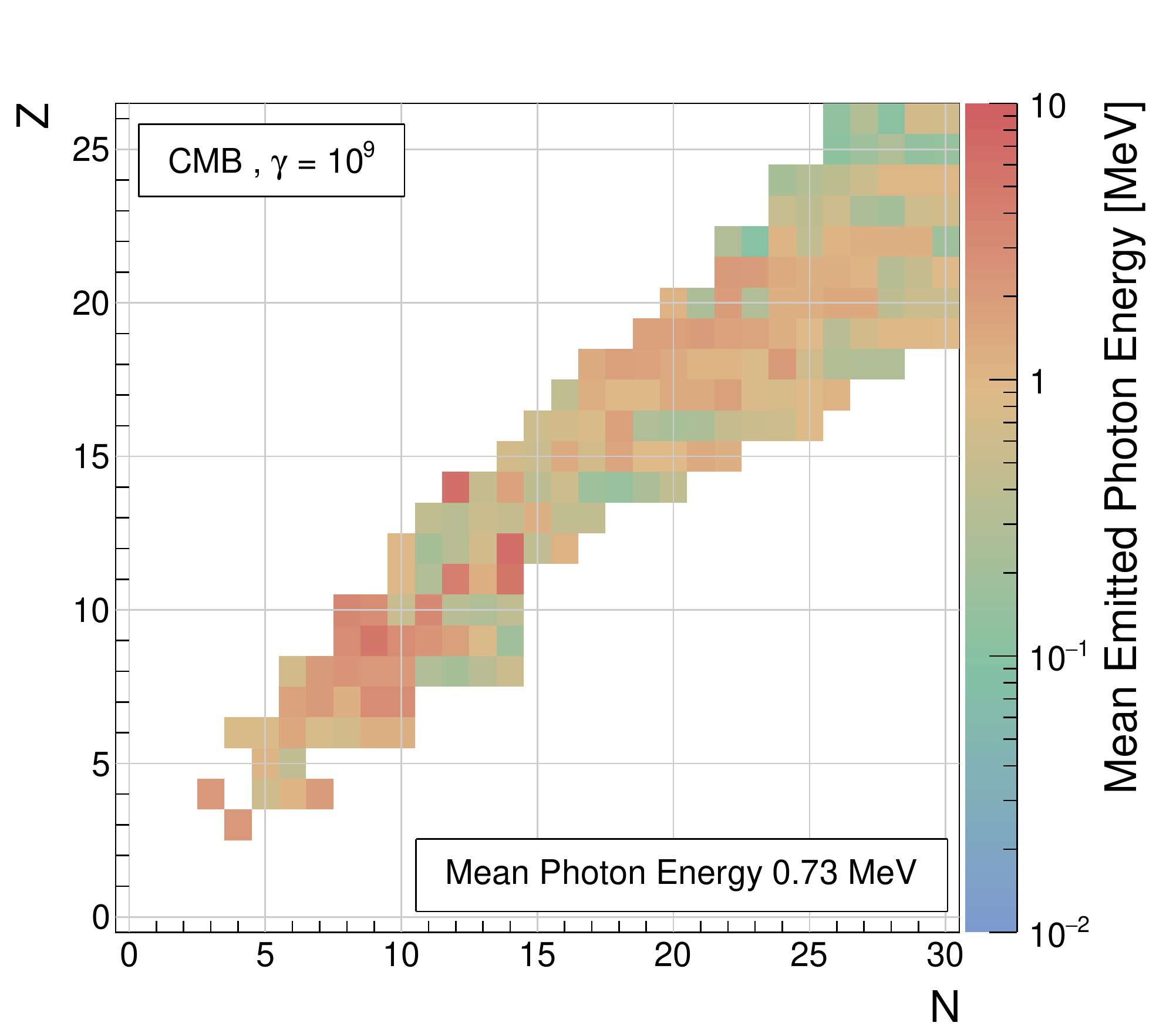}
\caption{
	Chart of nuclides considered for photon emission after photodisintegration.
	Color-coded is the mean emitted photon energy in the nucleus rest frame for an interaction with the CMB at a Lorentz factor of $\gamma = \num{e9}$.
	The emitted photon energy averaged over all nuclides, \SI{0.73}{MeV}, corresponds to PeV energies in the observer frame.
}
\label{fig:pd-mean-energy}
\end{center}
\end{figure}

For illustration we analyze the mean emitted photon energy for each of the 178 considered mother nuclei interacting with the CMB at a Lorentz factor of $\gamma = \num{e9}$.
The emitted photon energies of each daughter nucleus of a mother nucleus are weighted with the corresponding branching ratios, while for simplification the branching ratios for the disintegration to the individual daughter nuclei are assumed to be equal.
The resulting mean emitted photon energies are depicted in Figure \ref{fig:pd-mean-energy}.
On average, the emitted photon energy in the nucleus rest frame is approximately \SI{0.7}{MeV}.
Therefore, the observable energies of photons created via PD are expected to be in the PeV energy range.

\subsection{Elastic scattering}
\label{sec:production-scattering}

At energies below the photodisintegration threshold, background photons elastically scatter off cosmic-ray nuclei, $\ce{^A_ZX} + \gamma_\mathrm{bg} \longrightarrow \ce{^A_ZX} + \gamma$.
No additional particles are produced in this process.
However, the interacting background photon effectively receives a double Lorentz boost, accelerating it to cosmic-ray energies.

\begin{figure}[htbp]
\begin{center}
\includegraphics[width=0.47\textwidth]{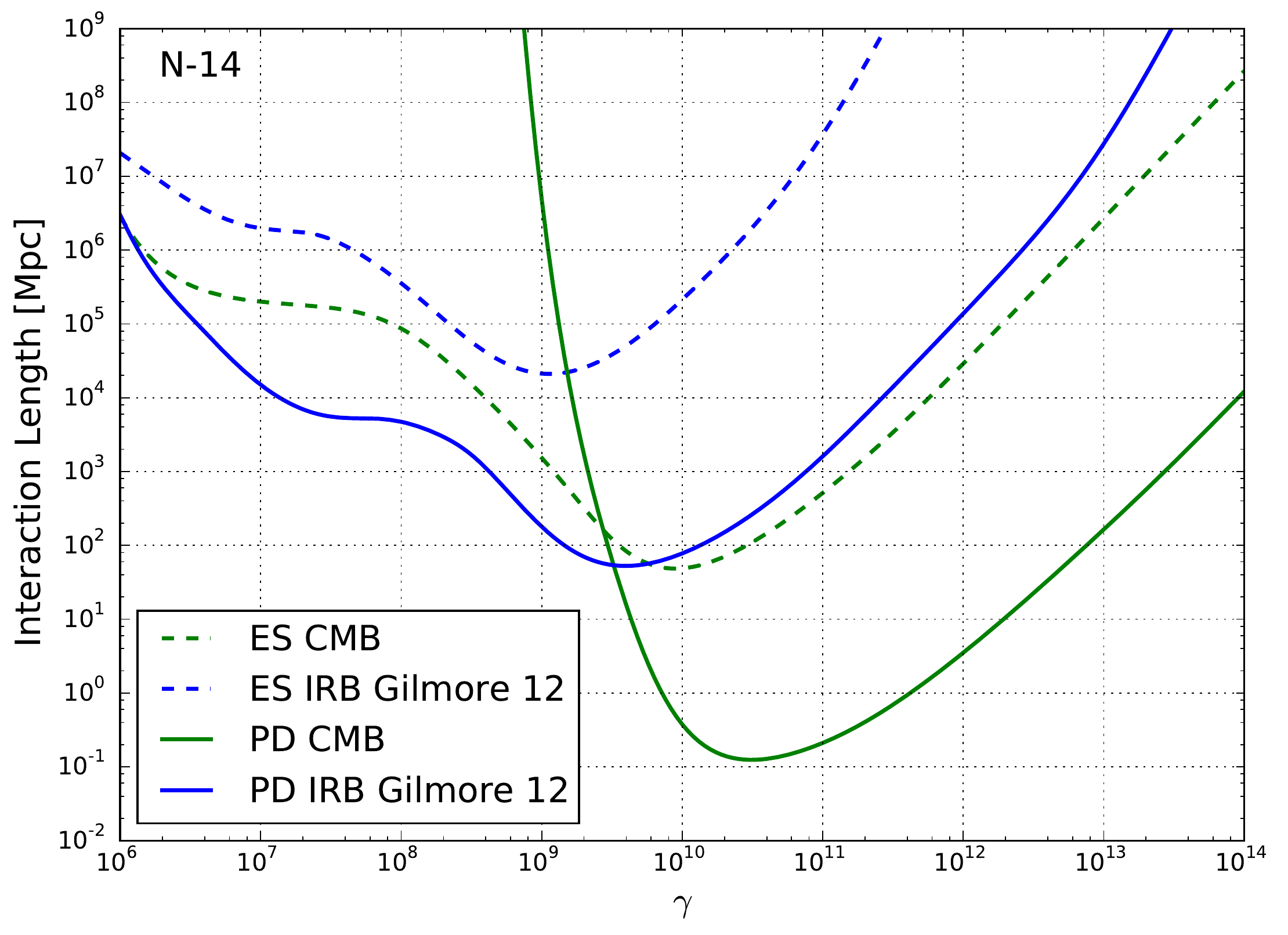}
\caption{
	Interaction length for elastic scattering (ES, dashed lines) in comparison to the interaction length for photodisintegration (PD, solid lines) for the case of \ce{^{14}N}.
	The interaction lengths are calculated for the CMB (green) and the IRB (blue).
}
\label{fig:PD_ES_interaction_length}
\end{center}
\end{figure}

Elastic scattering is implemented in CRPropa 3.1 as new photon production channel via a separate module.
The corresponding cross-sections are simulated using TALYS 1.8 \cite{TALYS,TALYS18} as described in the previous section.
The resulting interaction rates are shown in Figure \ref{fig:PD_ES_interaction_length} for the example of \ce{^{14}N}.
The interaction rates for elastic scattering are smaller by at least one order of magnitude compared to that for PD, except at the turnover where PD occurs dominantly on the CMB and IRB, respectively.
Interestingly, since elastic scattering requires only low photon energies in the nucleus rest frame, CMB photons are more likely than IRB photons to participate at all cosmic-ray energies.

\begin{figure*}[t]
\begin{center}
\begin{tabular}{cc}
	\parbox[c]{0.48\textwidth}{\includegraphics[width=0.48\textwidth]{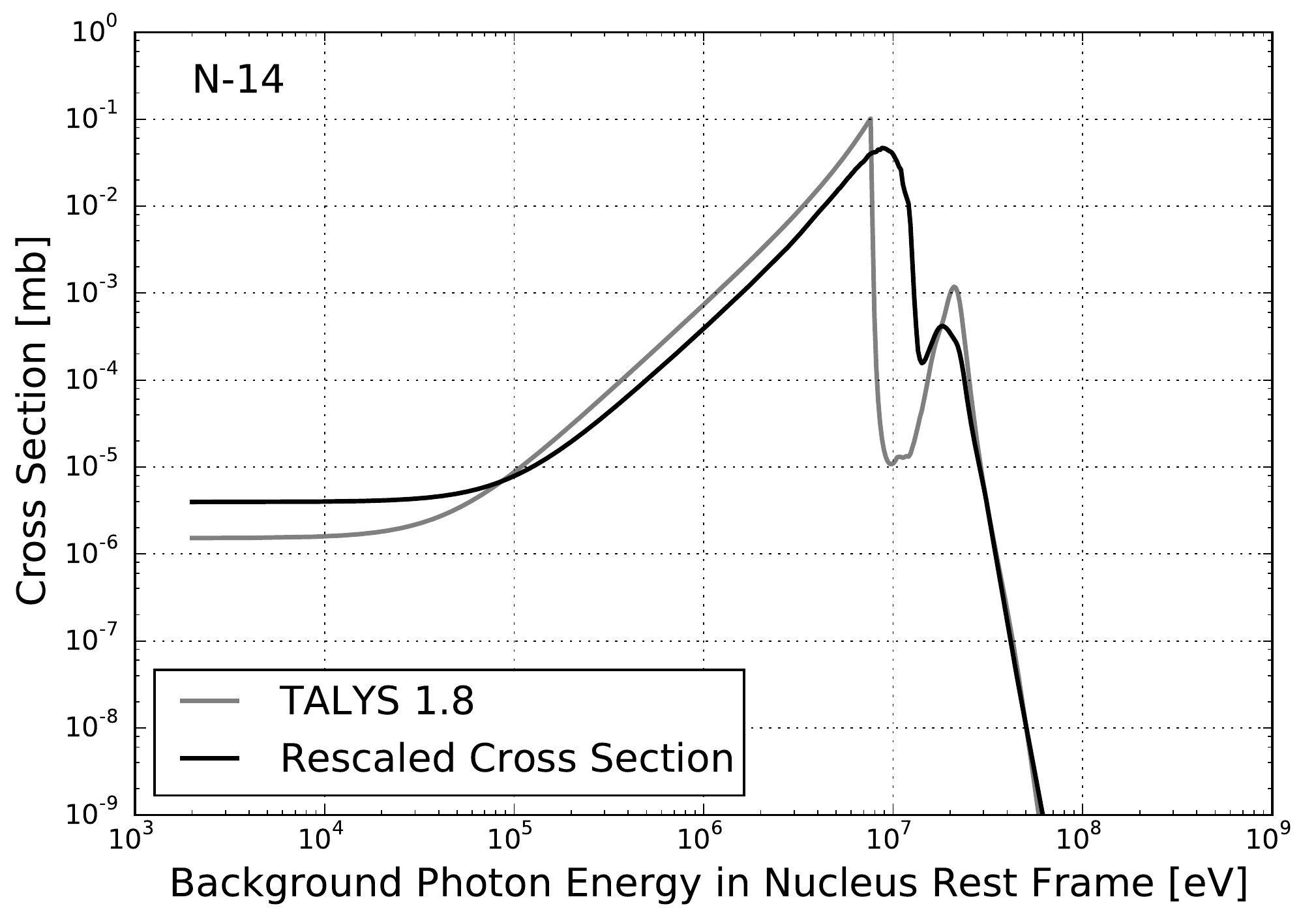}}
	&
	\parbox[c]{0.48\textwidth}{\includegraphics[width=0.48\textwidth]{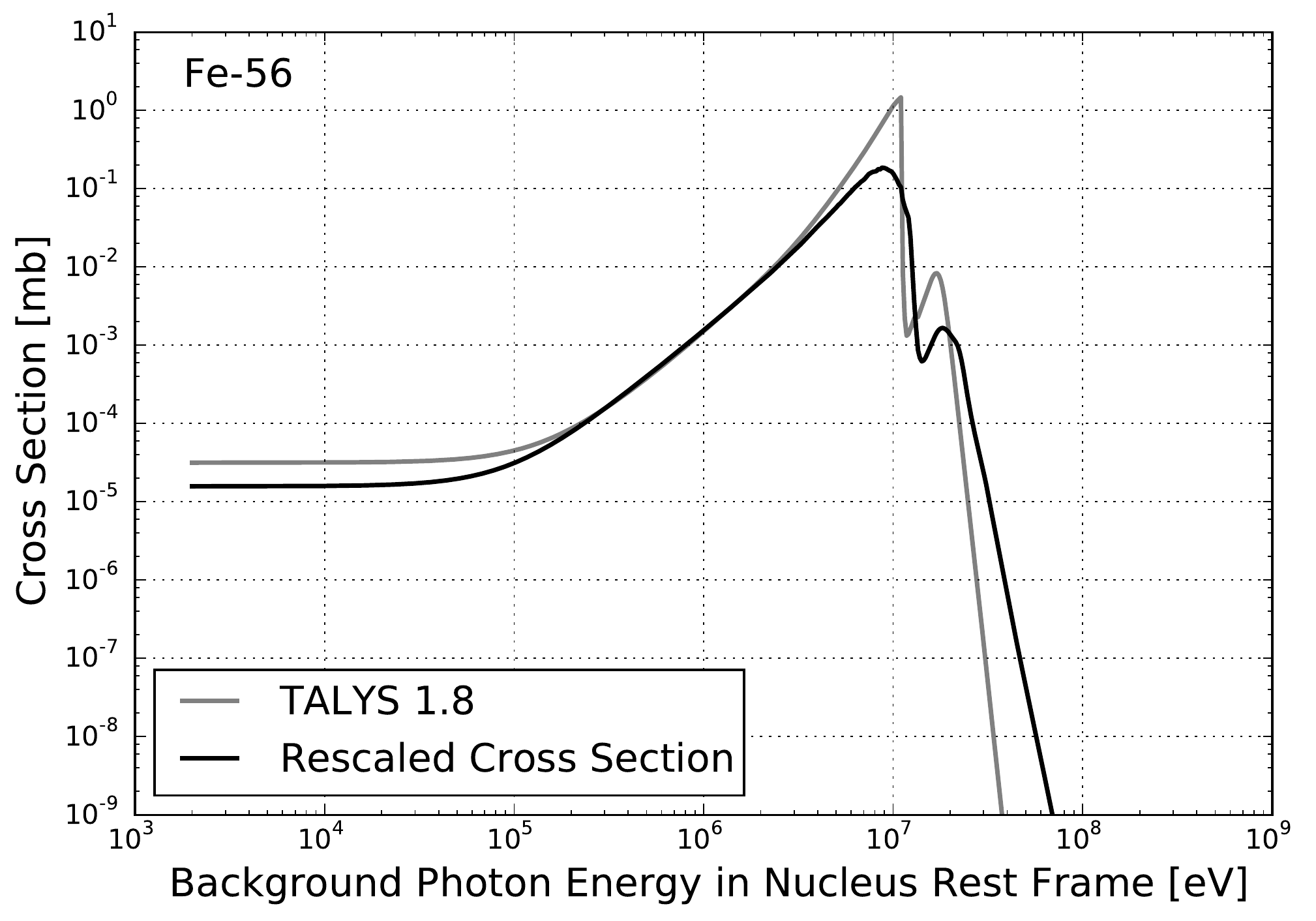}}
\end{tabular}
\caption{
	Comparison between original (gray) and approximate (black) cross-section for elastic scattering of a \ce{^{14}N} (left) and \ce{^{56}Fe} (right) nucleus.
	Above $\sim \SI{10}{MeV}$ the photodisintegration cross-section dominates.
	In the relevant region from \num{0.1} to \SI{10}{MeV} the cross-sections are in good agreement.
}
\label{fig:es-scaling}
\end{center}
\end{figure*}

In order to reduce the size of tabulated data for this subdominant process, the following approximation is applied.
According to the Thomas-Reiche-Kuhn sum rule in nuclei \cite{Rachen} the total integrated photoabsorption cross-section scales with the nuclear mass $A$ and charge number $Z$,
\begin{align*}
	\displaystyle\int_0^\infty \sigma(\epsilon) \, d\epsilon \approx \frac{2\pi^2 e^2 \hbar}{m_\mathrm{N}c} \frac{Z N}{A} = \sigma_0 \frac{Z N}{A}~.
\end{align*}
We factor out this principle $Z N / A$ scaling from the computed elastic scattering cross-sections and average them over all $n$ considered nuclides,
\begin{align*}
	\sigma_{\mathrm{avg}}(\epsilon) = \frac{1}{n} \sum_{Z,A} \frac{A}{ZN} \sigma_{Z,A}(\epsilon)~.
\end{align*}
An approximate cross-section for an individual nuclide can then be obtained by rescaling the average cross-section, $\sigma_{Z,A} \approx Z N \sigma_\mathrm{avg} / A$.
A comparison between the original and the approximate cross-sections is shown in Figure \ref{fig:es-scaling} for the examples of \ce{^{14}N} and \ce{^{56}Fe}.
The approximation is found to be in good agreement in the most relevant energy range from 0.1 to \SI{10}{MeV} in the nucleus rest frame.
Note that differences above the threshold of $\sim \SI{10}{MeV}$ are less relevant, since photodisintegration dominates here.

The scaling factors out when calculating the interaction rate on a given photon background field using equation (\ref{eq:rate}).
Hence we can write
\begin{align*}
	\lambda^{-1}_{Z,A}(E) = \frac{ZN}{A} \lambda^{-1}_\mathrm{avg}(E)~,
\end{align*}
which allows us to tabulate only $\lambda^{-1}_\mathrm{avg}$ and to apply the scaling during the simulation.
In case of an interaction, the energy of the interacting background photon $\epsilon$ is sampled from the differential interaction rate, cf. equation (\ref{eq:rate-differential}).
To this end, $d\lambda^{-1}_\mathrm{avg}/d\epsilon(\epsilon,\gamma)$ is tabulated for Lorentz factors $\gamma = \num{e6} - \num{e14}$ with 25 log-spaced points per decade, and photon energies $\epsilon = 0.002 - \SI{200}{MeV}$ with 100 log-spaced points per decade.
The photon is then boosted to the observer frame.

The scaling approximation has been validated by a one-dimensional simulation of \num{e4} nuclei with energies of \SI{50}{EeV}, source distances from 3 to \SI{1000}{Mpc} and considering all implemented interaction processes.
In order to provide a lower and an upper bound to the described approximation we perform the simulation three times: using the approximation, a lower bound and an upper bound to the differential interaction rate over all nuclides.
The resulting uncertainty on the emitted photon energies is small compared to the general simulation uncertainties.
All energy budgets are in the same order of magnitude, with the difference between the energy budget of the selected average and the two extreme cases being less than a factor of two.
Since the energy budget for elastic scattering is rather small, cf. section \ref{sec:results-budget}, the impact of this uncertainty is negligible in this context.

\subsection{Radiative decay}
\label{sec:production-decay}

Apart from stable elements, cosmic rays can also contain radioactive nuclides.
The implementation of nuclear decays is described in \cite{CRPropa2} and \cite{CRPropa3_Paper}.
This includes the emission of decay products $p, n, \alpha, e^\pm$ and $\nu_e$.
Here, we introduce the production of high-energy photons as nuclear decay products.

\begin{figure*}[t]
\begin{center}
\begin{tabular}{cc}
	\parbox[c]{0.48\textwidth}{\includegraphics[width=0.48\textwidth]{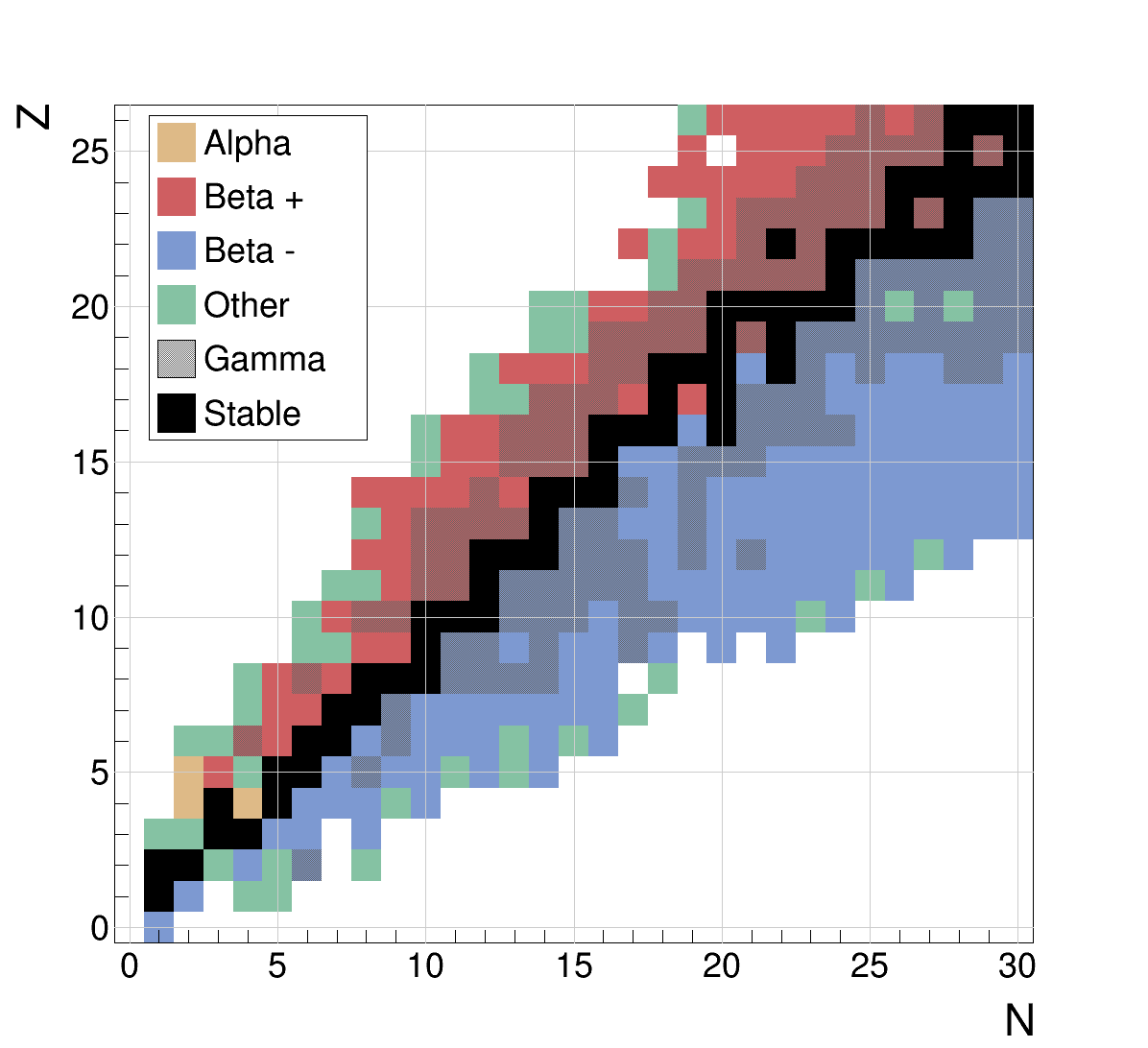}}
	&
	\parbox[c]{0.48\textwidth}{\includegraphics[width=0.48\textwidth]{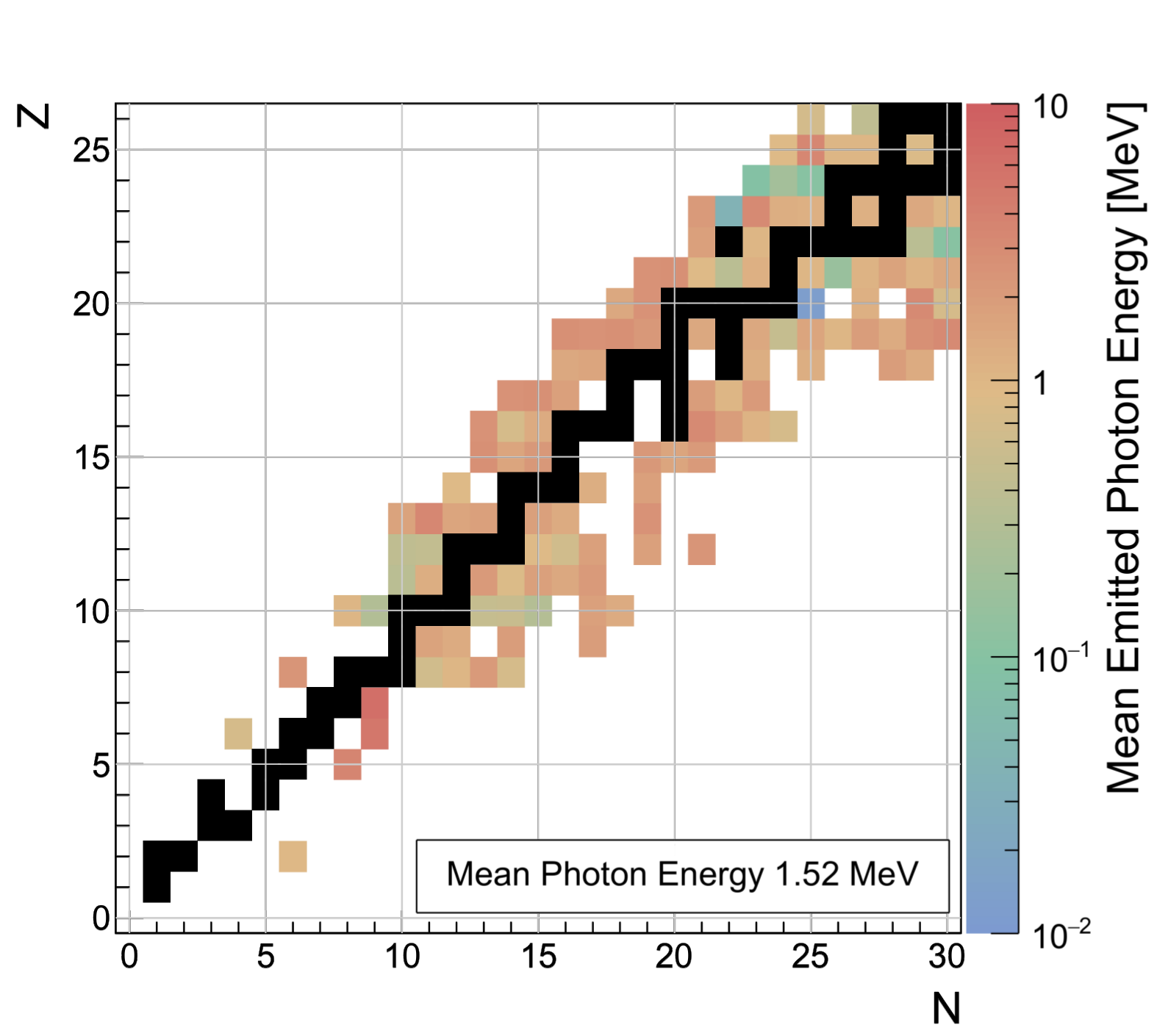}}
\end{tabular}
\caption{
	Nuclear decay tables in CRPropa.
	(Left) Dominant decay type for each nuclide.
	Nuclides with subsequent $\gamma$ decays are shaded gray.
	(Right) Mean emitted photon energy per $\gamma$ decay.
	The emitted photon energy averaged over all nuclides, \SI{1.52}{MeV}, corresponds to PeV energies in the observer frame at a Lorentz factor of \num{e9}.
}
\label{fig:decay}
\end{center}
\end{figure*}

One or more photons can be emitted in the nuclear relaxation process following a radioactive decay, $\ce{^A_Z X^*} \rightarrow \ce{^A_Z X} + n\gamma$.
The number and energy of the emitted photons depend on the preceding decay, as well as on nuclear level transitions inside the excited nucleus.
We use the experimental data provided by the NuDat 2.6 \cite{NuDat} database, which contains information for a total of 124 $\gamma$ decay channels.
For each $\gamma$ decay, the initial nuclide and preceding decay type is extracted, together with the photon energy and associated emission probability.
The NuDat decay radiation search also contains data from other processes and is therefore parsed to extract the relevant information.
Since cosmic-ray nuclei are completely ionized, the production of conversion electrons, Auger electrons and photons from positron annihilation are discarded.
Isomeric transitions from long-lived metastable nuclear levels are not considered, as this process is combined with electron shell processes in the NuDat database.
A nuclide chart with the dominant decay modes is shown on the left of Figure \ref{fig:decay}.
Nuclides with possible subsequent $\gamma$ decays are shaded gray.

In case of a radioactive decay during a simulation, subsequent $\gamma$ decays are performed randomly based on the according emission probabilities.
In order to ensure energy conservation, all emitted photon energies are taken into account when calculating the energies of the secondary particles created in the preceding decay.
All secondary particles are then boosted to the observer frame.

To illustrate the photon energies expected from $\gamma$ decays, on the right of Figure \ref{fig:decay} we depict the mean emitted photon energy for each nuclide.
The average over all nuclides is $\SI{1.5}{MeV}$.
Accordingly, for particles with a Lorentz factor of $\num{e9}$ an observable photon energy in the PeV range is expected.\section{Propagation of EM particles}
\label{sec:propagation}

In this section we describe the new implementations in CRPropa 3.1 for simulating the propagation of electromagnetic particles.
While propagating through intergalactic space, cosmic-ray photons experience (double) pair production
\begin{itemize}
	\item pair production (PP): \\ $\gamma + \gamma_\mathrm{bg} \longrightarrow e^+ + e^-$
	\item double pair production (DPP): \\ $\gamma + \gamma_\mathrm{bg} \longrightarrow e^+ + e^- + e^+ + e^-$
\end{itemize}
whereas cosmic-ray electrons and positrons (summarized as electrons hereafter) experience triplet pair production and inverse Compton scattering
\begin{itemize}
	\item triplet pair production (TPP): \\ $e + \gamma_\mathrm{bg} \longrightarrow e + e^+ + e^-$
	\item inverse Compton scattering (ICS): \\ $e + \gamma_\mathrm{bg} \longrightarrow e + \gamma$~,
\end{itemize}
by interacting with low energy background photons $\gamma_\mathrm{bg}$.
Electrons, being light charged particles, can also suffer significant energy losses via synchrotron radiation in the presence of magnetic fields.
The secondary particles from these processes interact in turn, forming an electromagnetic cascade down to GeV energies.

\begin{figure*}[t]
\begin{center}
\begin{tabular}{cc}
\parbox[c]{0.48\textwidth}{\includegraphics[width=0.48\textwidth]{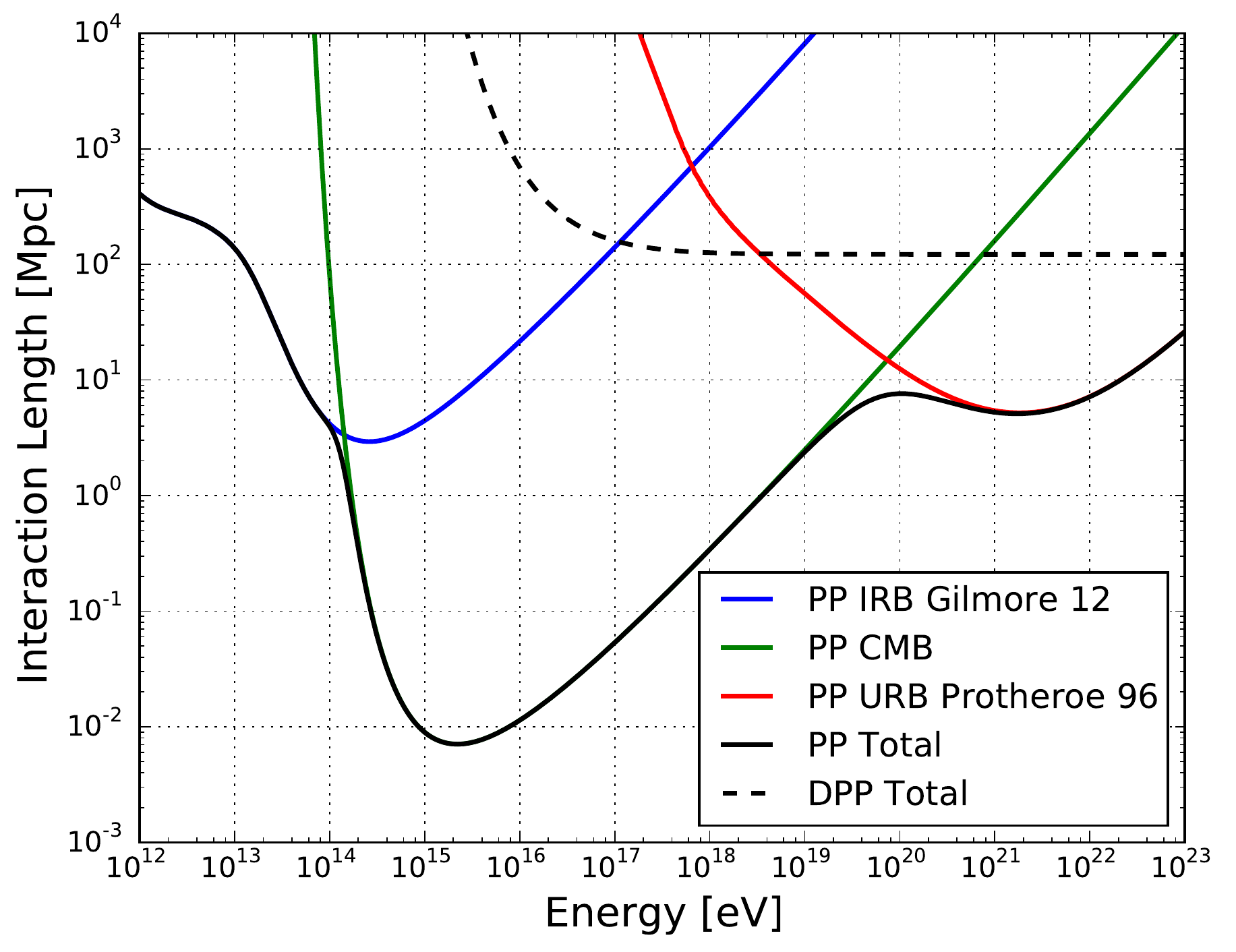}}
&
\parbox[c]{0.48\textwidth}{\includegraphics[width=0.48\textwidth]{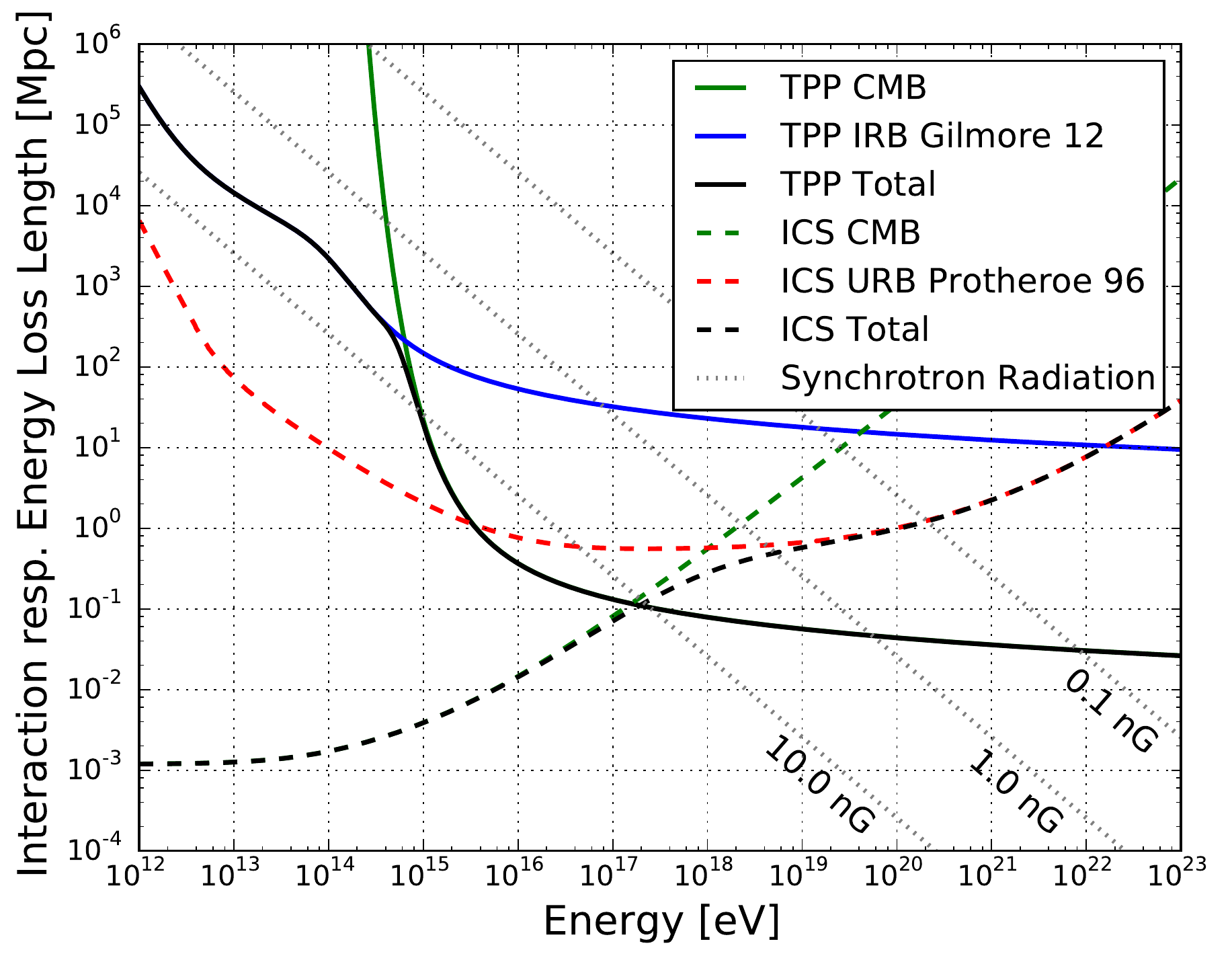}}
\end{tabular}
\caption{
	Interaction lengths for cosmic-ray photons (left) and electrons (right) interacting with cosmic photon backgrounds.
	The relevant processes are pair production (PP, black solid line) and double pair production (DPP, black dashed line) for photons, and triplet pair production (TPP, black solid line) and inverse Compton scattering (ICS, black dashed line) for electrons.
	Colored lines show the contribution of the individual photon fields, IRB (blue, Gilmore model~\cite{Gilmore2012}), CMB (green) and radio background (red, Protheroe model~\cite{Protheroe1996}).
	In addition, the energy loss length of synchrotron radiation is indicated for three magnetic field strengths (gray dotted lines).
}
\label{fig:em-interaction-length}
\end{center}
\end{figure*}

The importance of the individual processes can be assessed by comparing the interaction lengths $\lambda$, see Figure \ref{fig:em-interaction-length}.
For cosmic-ray photons, pair production is the most important interaction process at ultra-high energies.
Double pair production is rather sub-dominant and becomes relevant only at energies above \SI{e20}{eV}.
For cosmic-ray electrons, inverse Compton scattering dominates the interaction below $\sim \SI{e17}{eV}$, while above this energy triplet pair production is most relevant.
Synchrotron radiation can dominate at the highest energies depending on the magnetic field strength.

In the following we first describe the basic concepts of electromagnetic cascades in CRPropa before discussing the implementation of each process.

\subsection{General implementation}
\label{sec:propagation-general}

Following the general design of the CRPropa framework, each of the aforementioned processes is implemented in CRPropa 3.1 as an independent simulation module.
In contrast, in CRPropa 3.0 \cite{CRPropa3_Paper} the EleCa code \cite{EleCa} was used for this purpose as an external program: photons and electrons produced during the propagation of cosmic ray hadrons were collected and passed to EleCa once the main simulation was finished.
The present implementation fully replaces the EleCa code, thereby allowing the simulation of the propagation of photons and electrons as an integral part of CRPropa with several advantages:
First, users can apply the same extensive infrastructure that is already available in CRPropa for simulating nuclei primaries to specify source properties, simulation environment and observer configurations.
Second, CRPropa directly improves the electromagnetic interaction processes through the availability of more photon background models and of structured magnetic fields for synchrotron radiation and three-dimensional propagation.
Third, the general advantages of a modular simulation layout apply: a simulation can be set up using any combination of simulation modules, thereby enabling studies of individual interaction processes without modifications to the code.

While the new implementation largely follows that of EleCa, we have improved on a number of issues.
Most importantly the target photon energies are now correctly sampled from the individual photon backgrounds, which significantly modifies the spectrum of secondary particles \cite{master-heiter}.
The CRPropa interface to EleCa is currently still available for the purpose of comparisons, but will be removed in the next major release.

All processes for electromagnetic particles are implemented for cosmic-ray energies $E = \num{e10} - \SI{e23}{eV}$.
However, computational constraints currently limit the energy down to which the electromagnetic cascade can be simulated.
To illustrate, a single particle at EeV energy can result in \num{e6} particles at TeV energies, which increases the computing time accordingly.
Additionally, a memory requirement arises from the necessity to store all created secondaries until the propagation of the primary particle is completed.
To reduce this memory requirement, we have added a second mode of propagation in which secondary particles are propagated first before continuing with the primary particle.
This new mode is limited to energies above \SI{e15}{eV} for typical UHE scenarios.

For lower energies, the cosmic-ray transport code DINT \cite{DINT} can be used as an external program from within CRPropa, cf. \cite{CRPropa3_Paper} for details.
DINT propagates an entire cosmic-ray population in discrete steps in one dimension, hence the computation scales linearly with the energy range.
For one-dimensional simulations the DINT code can be combined with EleCa or CRPropa.
Here, particles above a specified crossover energy are simulated first using EleCa or CRPropa, before the remaining part of the cascade below the crossover energy is simulated with DINT.

\subsection{Pair production (Breit-Wheeler)}
\label{sec:propagation-pair}

The cross-section for electron pair production by photons (Breit-Wheeler) is given by \cite{DINT}
\begin{align*}
	\sigma_\mathrm{PP} =& \frac{3}{16}\; \sigma_\mathrm{T}\; (1-\beta^2)  \\
                            & \cdot \left[ (3-\beta^4) \ln\left( \frac{1+\beta}{1-\beta} \right) - 2 \beta (2-\beta^2) \right]~,
\end{align*}
with $\beta(s) = (1 - 4 m_e^2 c^4 s^{-1})^{1/2}$, $\sigma_\mathrm{T}$ the Thompson cross-section, and $s$ the squared center-of-mass energy.
With this cross-section, the interaction rate $\lambda^{-1}(E)$ is tabulated using equation (\ref{eq:rate}).
Likewise, the differential interaction rate $d\lambda^{-1}/ds\,(s,E)$ is tabulated using equation (\ref{eq:rate-differential}).
In case of an interaction, the cosmic-ray photon is consumed and an electron-positron pair with energies $E_{e^-},~E_{e^+}$ is created.
In order to determine $E_{e^-}$, first the squared center-of-mass energy $s$ of the interaction is sampled from $d\lambda^{-1}/ds\,(s,E)$.
Then, the ratio of photon to electron energy $x = E / E_{e^-}$ is sampled from the differential cross-section \cite{DINT}
\begin{align*}
\frac{d\sigma_\mathrm{PP}}{dx} = \frac{3}{4} \; \sigma_{\mathrm{T}} \; \frac{m_e^2 c^4}{s} \; \xi_\mathrm{PP}(x)
\end{align*}
with
\begin{align*}
\xi_\mathrm{PP}(x) =& \frac{x}{1-x} + \frac{1-x}{x}
          + (1-\beta^2)\left(\frac{1}{x}+\frac{1}{1-x}\right)\\
        & - \frac{1}{4}(1-\beta^2)^2 \left(\frac{1}{x}+\frac{1}{1-x}\right)^2 \;.
\end{align*}
Here $x$ is bounded by $\frac{1}{2}(1 - \beta) \leq x \leq \frac{1}{2}(1 + \beta)$.
From the sampled ratio $x$ the electron energy is readily calculated.
The positron energy $E_{e^+} = E - E_{e^-}$ then follows from energy conservation.
The small energy of the background photon can be neglected in this context.

\subsection{Double pair production}
\label{sec:propagation-doublepair}

We follow \cite{Brown_DPP} in approximating the double pair production cross-section as
\begin{align*}
	\sigma_\mathrm{DPP} = \SI{6.45}{\micro\barn}\left(1-\frac{16 m_e^2 c^4}{s} \right)^6~.
\end{align*}
The cross-section rapidly increases, starting from the minimum center-of-mass energy up to the final level of $\SI{6.45}{\micro \barn}$.
Since the pair production cross-section is much larger in the region where the double pair production cross-section is not constant, a more accurate parametrization of the double pair production's low energy part is not required.
In case of an interaction, the cosmic-ray photon is consumed.
As in \cite{DINT} we approximate that one of the electron-positron pairs receives all of the energy and that the energy is evenly distributed within this pair, $E_{e^-_1} = E_{e^+_1} = E / 2$ and $E_{e^-_2} = E_{e^+_2} = 0$.
Hence, no sampling procedure for the energy of the secondary particles is required.

\subsection{Triplet pair production}
\label{sec:propagation-tripletpair}

For triplet pair production we consider the cross-section from \cite{DINT}, reading
\begin{align*}
	\sigma_\mathrm{TPP} = \frac{3}{8}\; \sigma_\mathrm{T}\;\frac{\alpha}{\pi}\;\left[\frac{28}{9}\ln\left(\frac{s}{m_e^2c^4}\right)-\frac{218}{27}\right]~,
\end{align*}
where $\alpha$ denotes the fine structure constant.
While this expression is valid only for $s \gg m_e^2 c^4$, a separate treatment for lower $s$ is not required since inverse Compton scattering is dominant there, see Figure \ref{fig:em-interaction-length} (right).
Therefore, the cross-section is tabulated for values of $s \geq 13.4 m_e^2 c^4$ yielding only positive values.
In case of an interaction, the cosmic-ray electron loses a fraction of its initial energy $E$ to produce an electron-positron pair of energies $E_{e^-},~E_{e^+}$.
Here, the squared center-of-mass energy $s$ is sampled from the differential interaction rate $d\lambda^{-1}/ds\,(E,s)$.
This corresponds to a background photon energy in the observer frame of $\epsilon = (s - m_e^2 c^4) / 4 / E$ in the most likely case of a head-on collision.
The energies of the produced electron and positron $E_{e^\pm}$ are then approximated using \cite{Mastichiadis}
\begin{align*}
	E_{e^\pm} = \num{0.57} m_e c^2 \left(\frac{\epsilon}{m_e c^2}\right)^{-0.56} \left(\frac{E}{m_e c^2}\right)^{0.44}~.
\end{align*}
The energy loss of the interacting electron is thus $\Delta E = 2 E_{e^\pm}$.

\subsection{Inverse Compton scattering}
\label{sec:propagation-comptonscattering}

Cosmic-ray electrons can up-scatter background photons to cosmic-ray energies in inverse Compton scattering interactions.
For this process we consider the cross-section given in \cite{DINT}
\begin{align*}
	\sigma_\mathrm{ICS} = \frac{3}{8} \; \sigma_\mathrm{T} \; \frac{m_e^2 c^4}{s} \; \frac{1}{\beta} \; \xi_\circ
\end{align*}
with
\begin{align*}
\xi_\circ =& \frac{2}{\beta(1+\beta)}(2+2\beta-\beta^2-2\beta^3) \\
     & -\frac{1}{\beta^2}(2-3\beta^2-\beta^{3})\ln\left(\frac{1+\beta}{1-\beta}\right)\;.
\end{align*}
Here $\beta = (s-m_e^2c^4) / (s+m_e^2c^4)$ denotes the outgoing electron's velocity in the center-of-mass frame.
Analogously to the other interactions, first the squared center-of-mass energy $s$ is sampled from the differential interaction rate $d\lambda^{-1}/ds\,(E,s)$.
The ratio $x = E'/E$ of initial and final electron energy is then sampled from the differential cross-section
\begin{align*}
	\frac{d\sigma_\mathrm{ICS}}{dx} = \frac{3}{8}\; \sigma_\mathrm{T}\; \frac{m_e^2c^4}{s}\; \frac{1+\beta}{\beta}\;  \xi_\mathrm{ICS}(x)
\end{align*}
with
\begin{align*}
\xi_\mathrm{ICS}(x) =& x+\frac{1}{x}+\frac{2(1-\beta)}{\beta}\left(1-\frac{1}{x}\right)\\
                       & +\frac{(1-\beta)^2}{\beta^2}\left(1-\frac{1}{x}\right)^2
\end{align*}
in the range $(1-\beta)/(1+\beta) \leq x \leq 1$ \cite{DINT}.
The energy of the up-scattered background photon is then determined via energy conservation $E_\gamma = E_e - E'_e$, where its low initial energy is neglected.

\subsection{Synchrotron radiation}
\label{sec:propagation-synchrotron}

Relativistic charged particles emit synchrotron radiation when they are deflected in magnetic fields.
This process is implemented as a continuous energy loss and is applied in every propagation step.
For an ultra-relativistic particle the energy loss per unit distance reads \cite{Jackson}
\begin{align*}
	-\frac{dE}{dx} = \frac{1}{6\pi\epsilon_0}\frac{e^2}{\rho^2}\gamma^4\beta^4~,
\end{align*}
where $\epsilon_0$ is the vacuum permittivity, $\rho = p / (q B_\perp)$ the gyroradius, $B_\perp$ the magnetic field strength perpendicular to the direction of flight and $q$, $p$, $\beta c$, $\gamma$ the particle's charge, momentum, velocity and Lorentz-factor, respectively.

\begin{figure*}[htbp]
\begin{center}
\begin{tabular}{cc}
\parbox[c]{0.48\textwidth}{\includegraphics[width=0.45\textwidth]{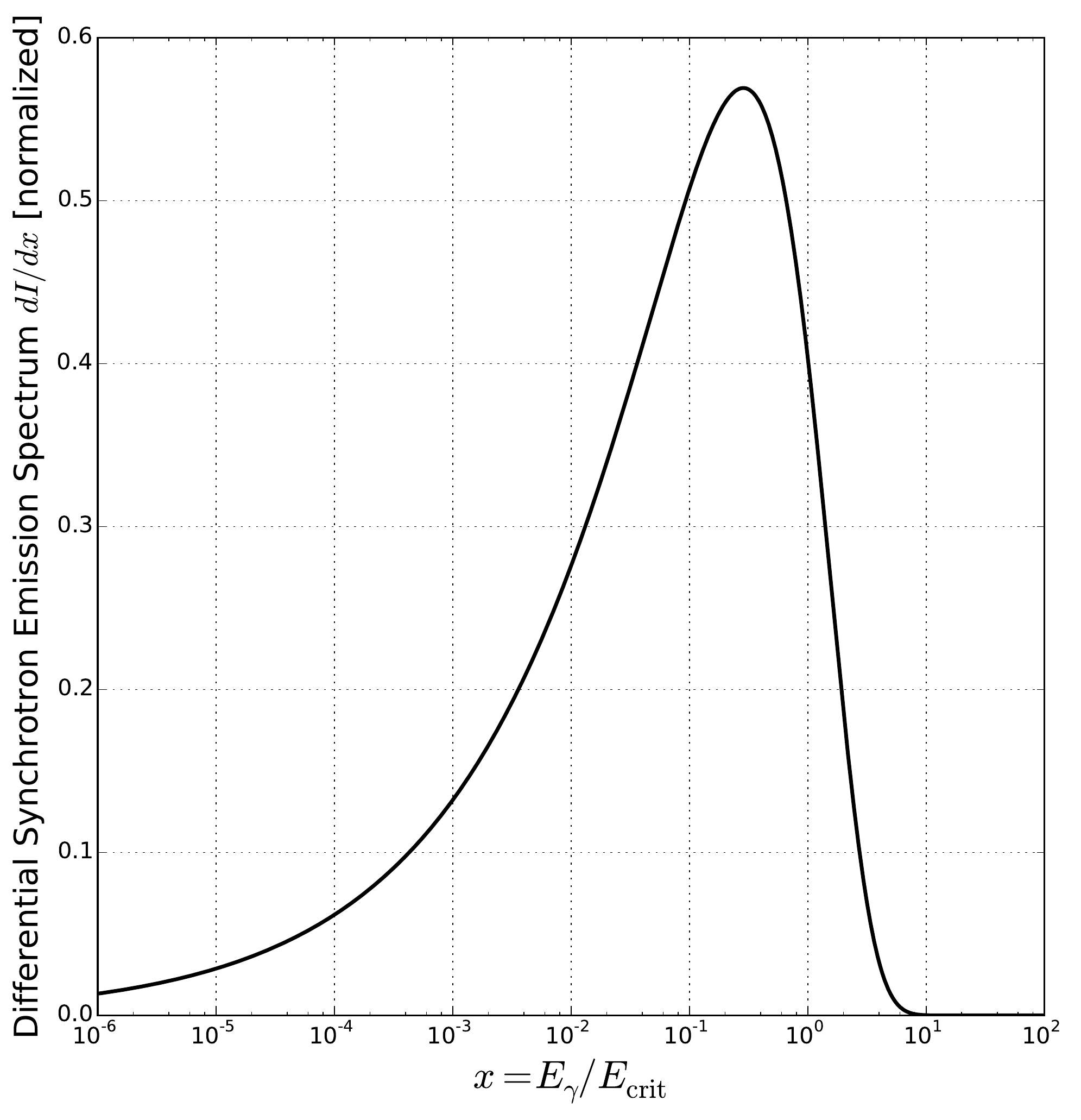}}
&
\parbox[c]{0.48\textwidth}{\includegraphics[width=0.48\textwidth]{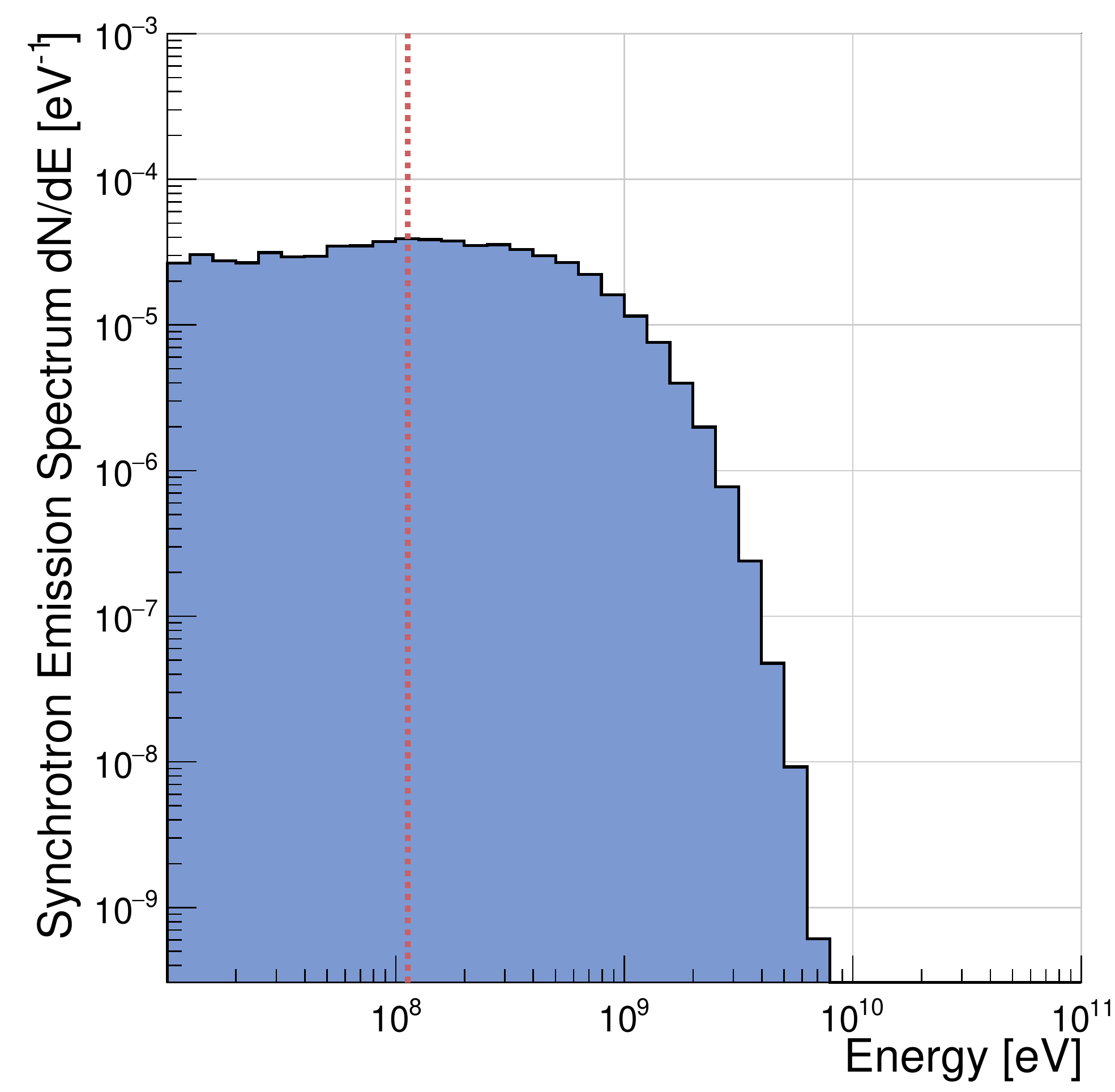}}
\end{tabular}
\caption{
	(Left) Spectrum of secondary photons from synchrotron radiation.
	(Right) Exemplary photon spectrum emitted by a \SI{10}{\exa\electronvolt} electron propagating over \SI{0.1}{kpc} in a magnetic field of \SI{0.1}{nG} RMS field strength.
	The red dashed line marks the most frequent energy.}
\label{fig:synchrotron}
\end{center}
\end{figure*}

The energies of the synchrotron photons are distributed according to
\begin{align*}
	\frac{dI}{dx} \propto x \int_x^\infty K_{5/3}(x')dx'~,
\end{align*}
where $K_{5/3}(x)$ is the modified Bessel function and $x = E_\gamma / E_\mathrm{crit}$ is the ratio of the synchrotron photon energy and the critical energy, given by $E_\mathrm{crit} = \frac{3}{2} \hbar c \gamma^3 / \rho$.
The distribution is shown in Figure \ref{fig:synchrotron} (left).
Secondary photons are sampled from this distribution until their total energy exceeds the energy loss of the primary particle in the current step.
Here, the last photon is randomly accepted to ensure energy conservation on average.
The synchrotron radiation module can be used in combination with a specific magnetic field, or via specifying the RMS field strength.
Due to the $\gamma^4$ dependency, synchrotron radiation is typically only relevant for UHE electrons.
For example, a $\SI{10}{\exa\electronvolt}$ electron loses about $\SI{0.3}{\exa\electronvolt\per Mpc}$ in a $\SI{0.1}{\nano G}$ magnetic field, whereas an iron nucleus of the same energy only loses about $\SI{2}{\electronvolt \per Mpc}$.

As an example, the differential energy spectrum of the generated synchrotron photons for a single \SI{10}{EeV} electron in a magnetic field with a RMS field strength of \SI{0.1}{nG} propagated over a distance of $\SI{0.1}{kpc}$ is shown in Figure \ref{fig:synchrotron} (right).
A large number of secondary photons is emitted even over this short distance.
\section{Results}
\label{sec:results}

\subsection{Comparison of photon production channels}
\label{sec:results-budget}

In the following we quantify the importance of the individual production channels for electromagnetic particles by cosmic-ray nuclei.
To this end, \num{e4} iron nuclei with energies of \SI{50}{EeV} emitted by uniformly distributed sources between \num{3} and \SI{1000}{Mpc} distance are simulated.
The cosmic rays are propagated in a one-dimensional simulation with all relevant interaction processes and using the CMB and the Gilmore~\cite{Gilmore2012} IRB model as photon backgrounds.
All photons and electrons are stored immediately after their production and are not propagated further.

\begin{figure}[ht!]
\includegraphics[width=0.5\textwidth]{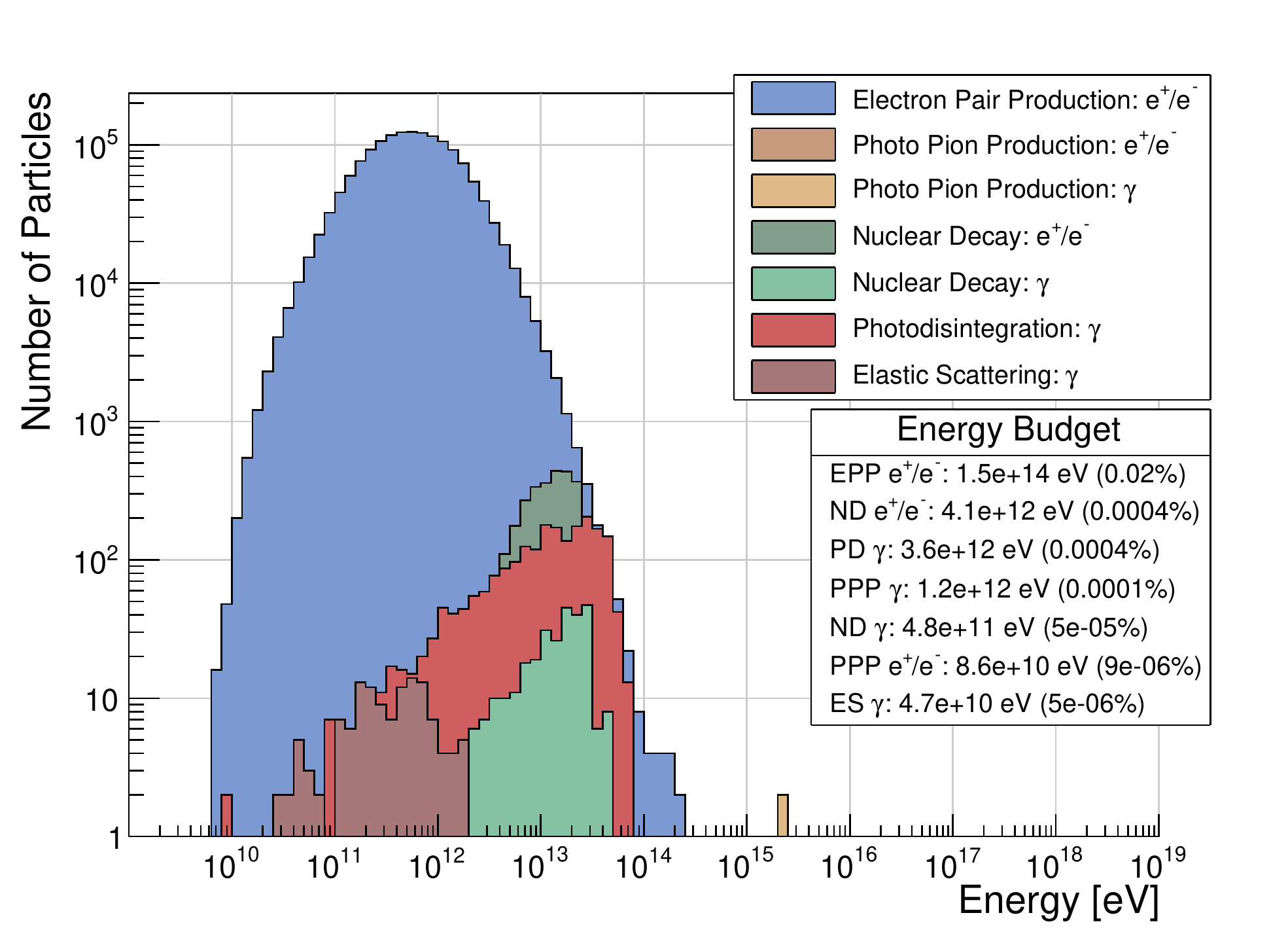}
\includegraphics[width=0.5\textwidth]{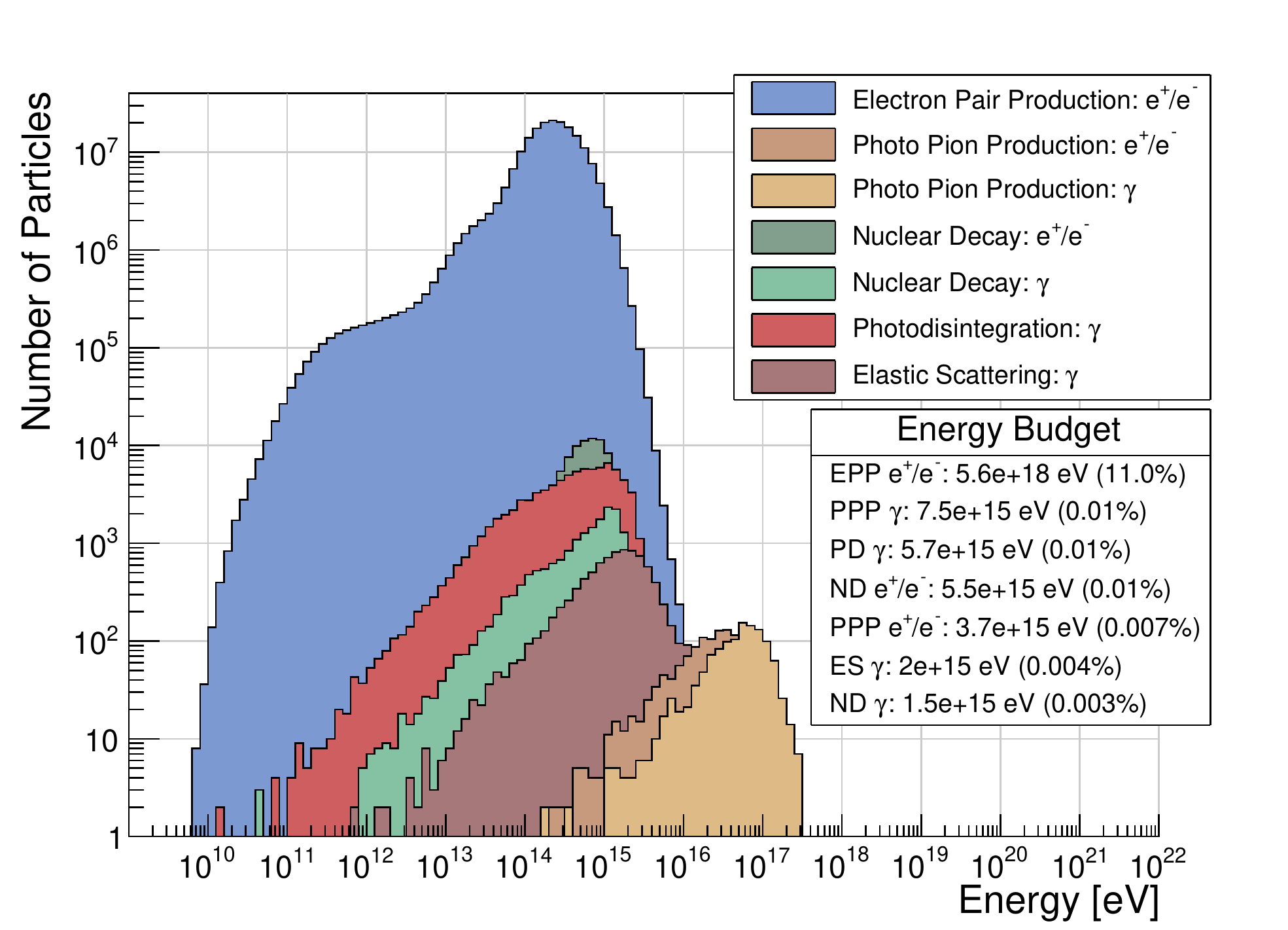}
\includegraphics[width=0.5\textwidth]{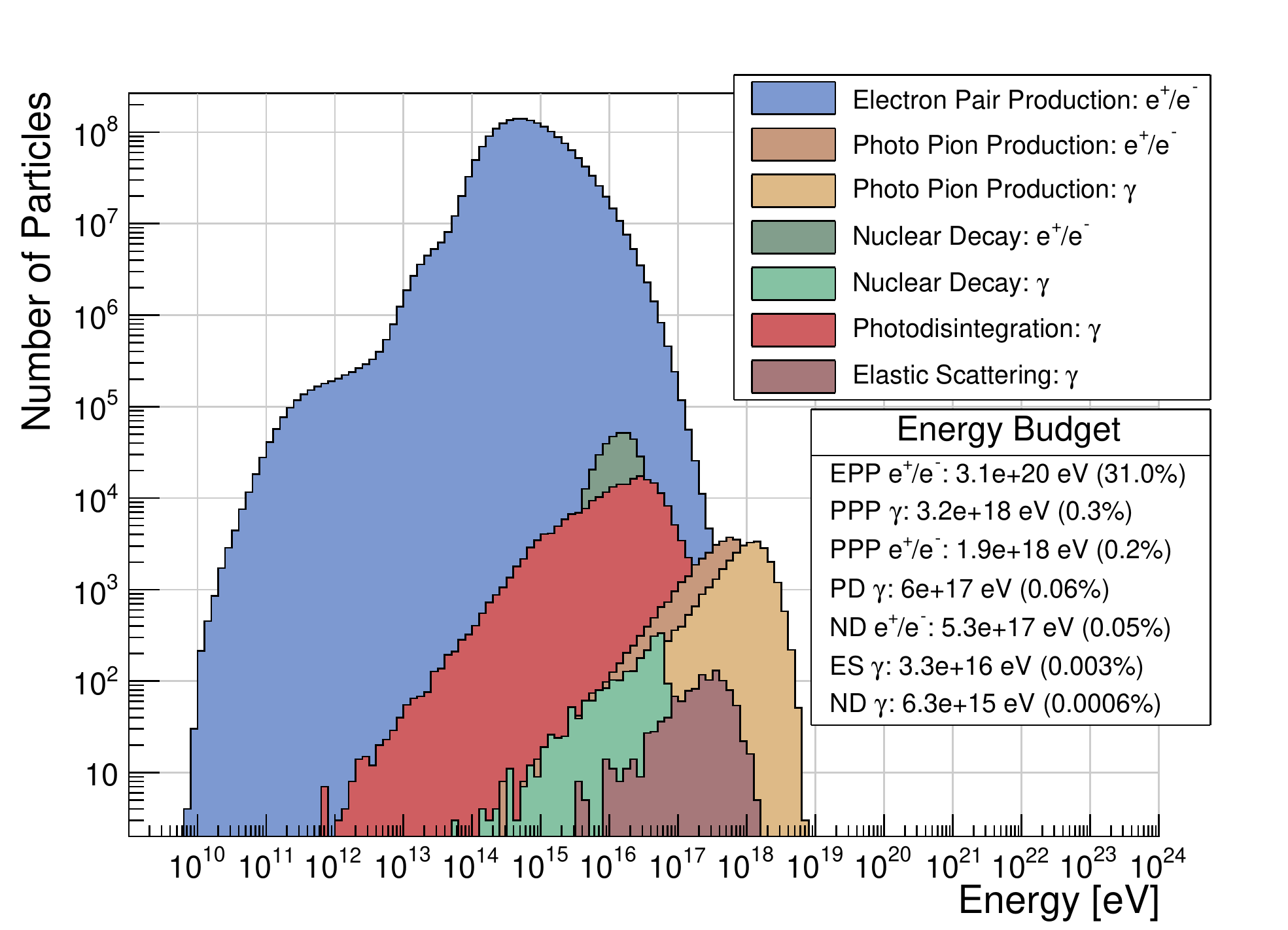}
\begin{center}
\caption{
Photon and electron emission spectra separated by photon production channel for \num{e4} iron nuclei with energies of \SI{1}{EeV} (top), \SI{50}{EeV} (center), \SI{1000}{EeV} (bottom) and source distances of $\num{3}-\SI{1000}{Mpc}$.
The energy budget of each production channel normalized to one iron nucleus is shown on the bottom right.
}
\label{fig:energy_budget}
\end{center}
\end{figure}

The resulting photon and electron emission spectra for the individual photon production channels are shown in the center of Figure \ref{fig:energy_budget}.
As can be seen, electron pair production forms the dominant contribution up to about \SI{e16}{eV}.
At higher energies, photons and electrons from photo-pion production dominate the spectrum up to the highest energies of about \SI{e17.5}{eV}.
Considering only photons, a remarkable contribution is generated via elastic scattering in the range from 2 to \SI{20}{PeV}.
For photon energies below \SI{2}{PeV} photodisintegration becomes dominant.
The photon emission spectra of photodisintegration and nuclear decay peak at about \SI{1}{PeV}.
This is in good agreement with the expected photon energies for these processes, cf.\ sections \ref{sec:production-disintegration} and \ref{sec:production-decay}, considering the Lorentz factor $\gamma \sim \num{e9}$ of a \SI{50}{EeV} iron nucleus.

The emission spectra strongly depend on the initial energy of the cosmic-ray nucleus.
In the top and bottom of Figure \ref{fig:energy_budget} we show for comparison the emission spectra for initial cosmic-ray nuclei energies of \SI{1}{EeV} and \SI{1000}{EeV}, respectively.
At \SI{1}{EeV} the contribution of the photo-pion production almost vanishes due to this processes' energy threshold, and the photodisintegration becomes the dominant photon production channel instead.
In contrast, at an extremely high energy of \SI{1000}{EeV} the importance of photo-pion production further increases with respect to photodisintegration and elastic scattering.

With the exception of elastic scattering, the energy of secondary particles in the nucleus rest frame depends little on the nucleus energy.
After boosting to the observer frame, the energy of secondary particles increases approximately linearly with the nucleus energy, which can be seen by the shift of the emission spectra for different initial energies.
In elastic scattering, the interacting background photon is Lorentz boosted twice; first to the nucleus rest frame and then back to the observer frame.
Hence, the observable photon energies scale approximately as the square of the nucleus initial energy.

The fraction of energy that is transferred from the cosmic-ray nucleus via the individual photon and electron production channels is compared in Table \ref{tab:energy_budget}.
These energy budgets are a measure of the importance of the individual production channels.
Overall, electron pair production is the most important process for transferring energy to the electromagnetic cascade.
With increasing energy of the primary nucleus, the fraction of energy transferred to electromagnetic particles increases as well.
All production channels contribute to this increase, except for nuclear decay and elastic scattering, whose contribution begins to decrease at the highest primary particle energies in the considered scenario.

\begin{table*}[t]
\centering
	\begin{tabular}{lcccc} \hline \hline
	Process & Spectrum & \multicolumn{3}{c}{Primary Particle Energy}  \\
	& & $\SI{1}{EeV}$ & $\SI{50}{EeV}$ & $\SI{1000}{EeV}$ \\ \hline
	Electron Pair Production & $e^{\pm}$ & $\SI{0.02}{\percent}$ & $\SI{11.0}{\percent}$ & $\SI{31.0}{\percent}$ \\
	Photo-pion Production & $e^{\pm}$ & $\SI{9e-6}{\percent}$ & $\SI{0.007}{\percent}$ & $\SI{0.2}{\percent}$ \\
	Photo-pion Production & $\gamma$ & $\SI{1e-4}{\percent}$ & $\SI{0.01}{\percent}$ & $\SI{0.3}{\percent}$ \\
	Nuclear Decay & $e^{\pm}$ & $\SI{4e-4}{\percent}$ & $\SI{0.01}{\percent}$ & $\SI{0.05}{\percent}$ \\
	Nuclear Decay & $\gamma$ & $\SI{5e-5}{\percent}$ & $\SI{0.003}{\percent}$ & $\SI{6e-4}{\percent}$ \\
	Photodisintegration & $\gamma$ & $\SI{4e-4}{\percent}$ & $\SI{0.01}{\percent}$ & $\SI{0.06}{\percent}$ \\
	Elastic Scattering & $\gamma$ & $\SI{5e-6}{\percent}$ & $\SI{0.004}{\percent}$ & $\SI{0.003}{\percent}$ \\ \hline \hline
\end{tabular}
\caption{
Energy budget of the individual photon and electron production channels for cosmic-ray iron emitted with 1, 50 and $\SI{1000}{EeV}$ from uniformly distributed sources between
$\num{3}$ and $\SI{1000}{Mpc}$ distance.
}
\label{tab:energy_budget}
\end{table*}

\subsection{Propagation of electromagnetic particles}
\label{sec:results-propagation}

In the following we assess the importance of the individual EM production channels in the observable photon flux.
We also compare the interaction processes for electromagnetic particles in CRPropa 3.1 with their implementation in the DINT \cite{DINT} code.
As case studies we consider the electromagnetic cascade induced by UHE cosmic-ray protons, respectively iron:
In both scenarios, \num{e4} cosmic rays are emitted with a power-law spectrum $dN/dE \propto E^{-1}$ between 1 and \SI{1000}{EeV} from uniformly distributed sources between 3 and \SI{1000}{Mpc} distance.
As before, the cosmic rays are propagated in a one-dimensional simulation considering all relevant interaction processes and using the CMB and the Gilmore~\cite{Gilmore2012} IRB model as photon backgrounds.

In order to factor out statistical fluctuations in the hadronic part of the simulation, the propagation of the iron nuclei is performed only once and the resulting secondary photons and electrons are saved directly at their point of creation.
The electromagnetic particles are then propagated with both simulation codes.
Here, CRPropa is used in combination with DINT as described in section \ref{sec:propagation-general} to extend the simulation down to \SI{e12}{eV}.
As crossover energy, we choose \SI{e17}{eV}, corresponding to the setup used in \cite{CRPropa3_Paper} for comparing DINT with the combination of DINT and EleCa.

\begin{figure*}[htb!]
\begin{center}
\begin{tabular}{cc}
\parbox[c]{0.48\textwidth}{\includegraphics[width=0.48\textwidth]{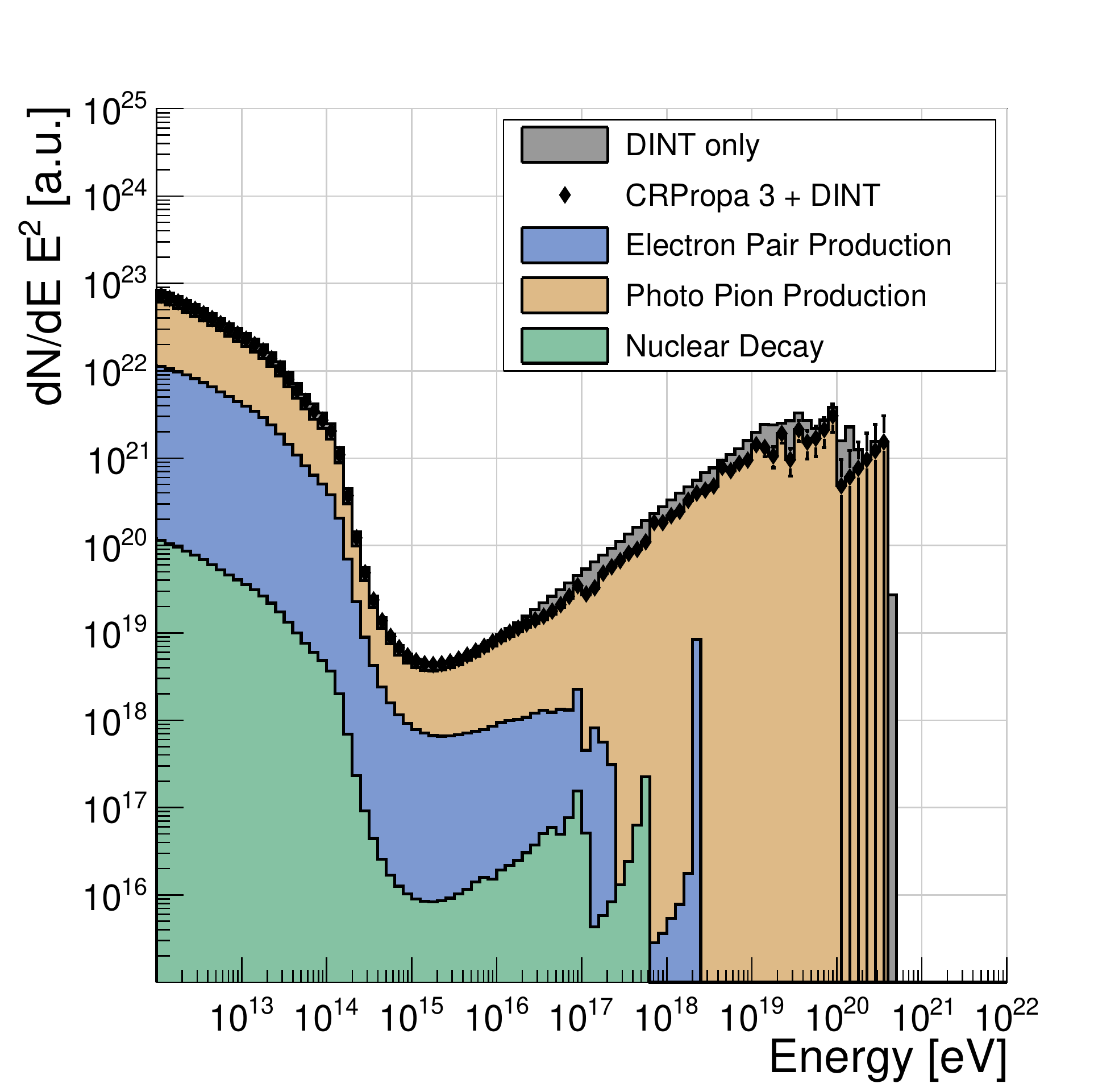}}
&
\parbox[c]{0.48\textwidth}{\includegraphics[width=0.48\textwidth]{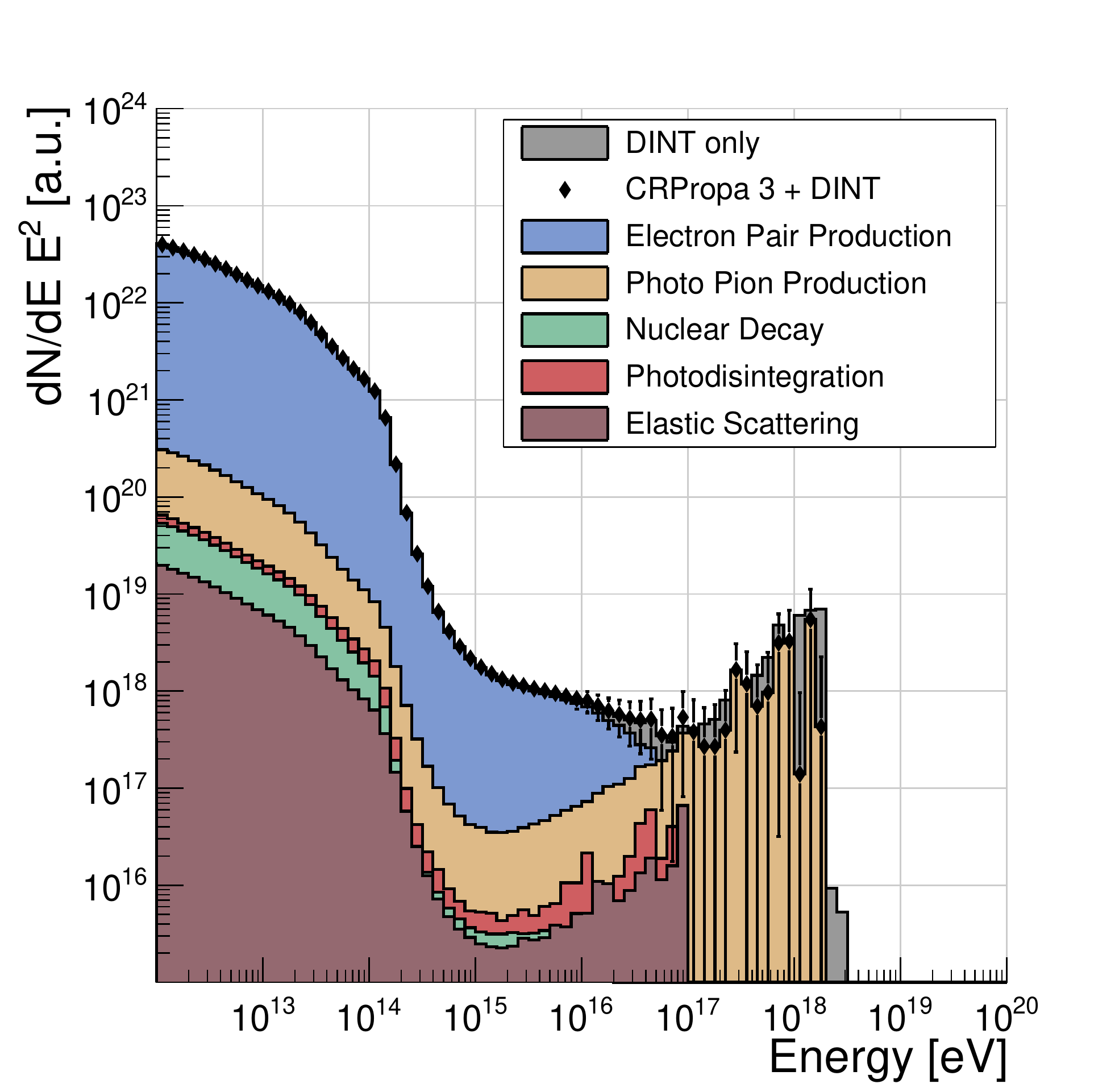}}
\end{tabular}
\caption{
Photon flux, scaled with $E^2$, created by \num{e4} cosmic rays (left: protons, right: iron) emitted with a power-law spectrum $dN/dE \propto E^{-1}$ between 1 and \SI{1000}{eV} from uniformly distributed sources at $\num{3}-\SI{1000}{Mpc}$ distance.
Black dots show the total photon flux simulated with CRPropa above \SI{e17}{eV} and DINT below, with error bars depicting the statistical uncertainty.
Color-coded are the contributions of the individual EM production channels.
The photon flux obtained from a pure DINT propagation is shown in gray for comparison.
}
\label{fig:flux_1e17}
\end{center}
\end{figure*}

The resulting photon fluxes in simulation are shown in Figure \ref{fig:flux_1e17} for the proton scenario (left) and the iron scenario (right).
In the iron scenario, the photon fluxes above \SI{e17}{eV} simulated with CRPropa (black dots) and DINT (gray curve) agree well, while in the proton scenario the flux simulated with CRPropa is slightly below that of DINT.

Additionally, the color-coded contributions of the individual production channels are shown for the combination of CRPropa and DINT.
In the iron scenario, photo-pion production marks the dominant contribution to the photon flux at highest energies, whereas electron pair production is most relevant for lower energies.
The newly implemented photon production channels via elastic scattering, photodisintegration and nuclear $\gamma$ decay are subdominant in this scenario.

In the proton scenario, the contribution of photo-pion production is dominant throughout the considered energy range.
The nuclear decay contribution originates from the $\beta$-decay of neutrons produced via photo-pion production.
Currently, protons are not considered for elastic scattering in CRPropa and there is thus no contribution of this channel.
Electron pair production and nuclear decay contribute photons subdominantly in this scenario.

\subsection{Calculating the UHE photon horizon}
\label{sec:results-horizon}

We now compute the UHE photon horizon as an example application of the EM propagation in CRPropa 3.1.
Typically, the photon horizon (or attenuation length) $D_{36.8\%}$ is defined as the distance at which the survival probability has dropped to $1/e = 36.8$~\%.
In addition, we also introduce the distance $D_{1\%}$ at which the survival probability has dropped to 1~\%.
In the electromagnetic cascade of a primary photon, further secondary photons are produced from high-energy electrons via inverse Compton scattering.
Since electrons are deflected by intervening magnetic fields, secondary photons from these electrons also exhibit an angular separation with respect to the primary photon.
For small deflections, UHE photon sources may also be detectable via secondary photons.
For this reason, we make use of the capability of CRPropa to track charged particles in various magnetic field models in order to compute the photon horizon including these secondary photons.

As a case study, which is motivated by the targeted photon search analysis presented in \cite{auger-targetedphotonsearch}, we additionally compute the photon horizon including secondary photons for energies above \SI{e17}{eV} and angular deflections smaller than \ang{1}.
In this three-dimensional simulation, sources at various distances isotropically emit \num{50000} mono-energetic photons.
As magnetic field model we consider a random turbulence with a Kolmogorov type power spectrum, a coherence length of \SI{1}{Mpc} and RMS field strengths of $B_{\rm RMS} = \SI{0.1}{\nano G}$ and $B_{\rm RMS} = \SI{1}{\nano G}$.
All interaction processes described in section \ref{sec:propagation} are considered, including the energy loss via synchrotron radiation and cosmological redshift.
As background photon fields, the CMB as well as the Gilmore~\cite{Gilmore2012} IRB model and the Protheroe~\cite{Protheroe1996} radio background model are considered.
The initial photons and all relevant secondary particles are propagated over a distance of \SI{50}{Mpc}.
The number of all photons with deflections below \ang{1} from the initial emission direction is recorded at fixed log-spaced distances between \SI{50}{kpc} and \SI{50}{Mpc}.
Interpolating between these values allows the calculatation of the photon horizon for a given survival probability.
The simulation is repeated for 61 log-spaced initial photon energies between \SI{0.1}{EeV} and \SI{100}{EeV}.

\begin{figure*}[t]
\begin{center}
\includegraphics[width=1.\textwidth]{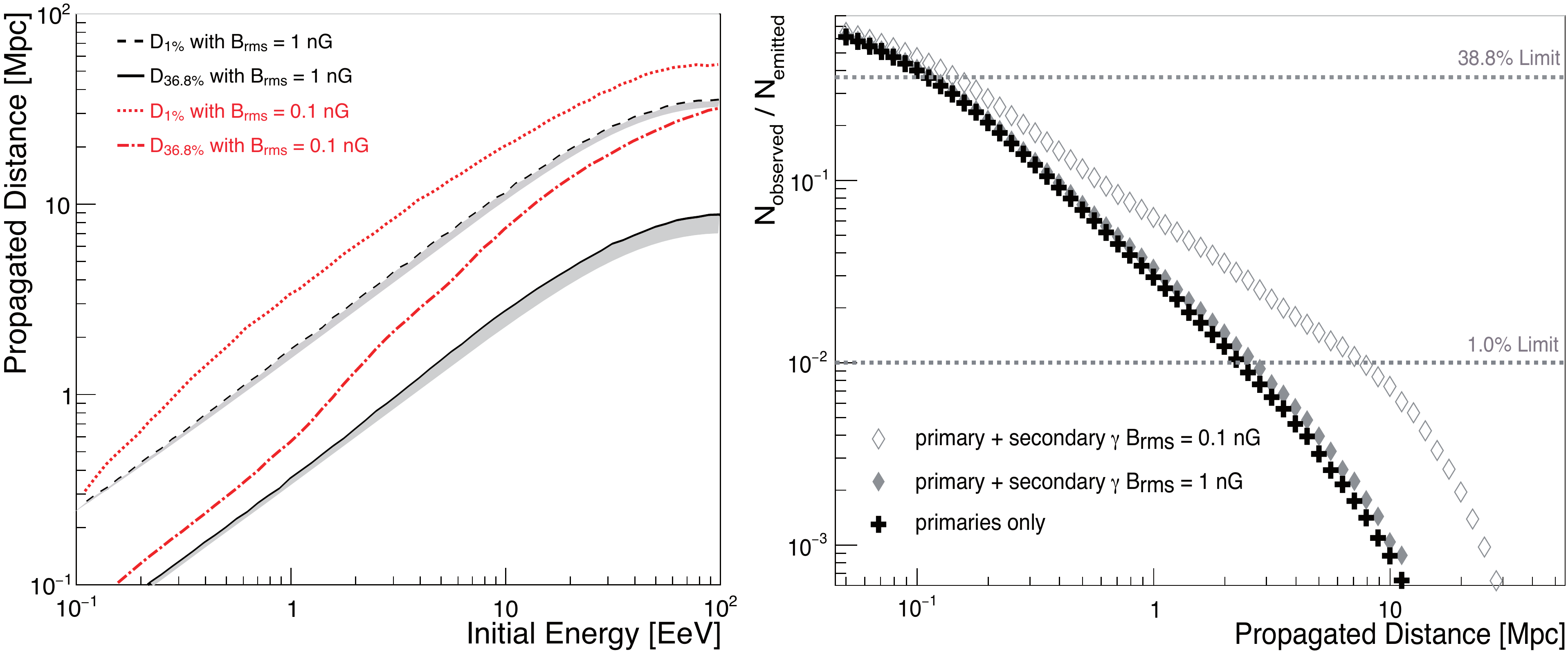}
\caption{
(Left) Photon horizon $D_{36.8\%}$ and $D_{1\%}$ including secondary photons with energies above \SI{e17}{eV} and deflections below \ang{1} considering a magnetic field strength $B_{\rm RMS}$ of \SI{0.1}{nG} (red) and \SI{1}{nG} (black). The gray shaded area indicates the difference of the horizon between primary and primary $+$ secondary photons for $B_{\rm RMS}$ = \SI{1}{nG}. (Right) Photon horizon for initial photon energies from $\num{0.1}-\SI{100}{EeV}$ according to a $dN/dE \propto E^{-2}$ source spectrum.
The ratio of photons observed at $E > \SI{e17}{eV}$ with angular deflections below \ang{1} to the emitted photons is shown as a function of distance.
}
\label{fig:horizon}
\end{center}
\end{figure*}

The resulting photon horizon including secondary photons is shown in Figure \ref{fig:horizon} (left). With $B_{\rm RMS} = \SI{0.1}{nG}$ (red lines) and an initial photon energy of \SI{10}{EeV} the horizon is about $D_{1\%} = \SI{20}{Mpc}$, which is twice as large as when considering only primary photons. For $D_{36.8\%}$ the horizon is about \SI{7}{Mpc}, which corresponds to an increase by a factor of more than three. At a ten times higher RMS magnetic field strength of \SI{1}{\nano G} (black lines) the increase of the horizon is much less pronounced, since secondary electrons are strongly deflected before they can produce further photons. This is indicated by the gray shaded area in Figure \ref{fig:horizon} (left) showing the difference between primary and primary $+$ secondary photons with $B_{\rm RMS} = \SI{1}{nG}$.

In addition to the aforementioned fixed initial photon energies, we compute the horizon for a photon source with a power-law spectrum $dN/dE \propto \sim E^{-2}$ between \num{0.1} and \SI{100}{EeV}. The results are shown in Figure \ref{fig:horizon} (right).
In this scenario, the photon horizon for the \SI{1}{\percent} limit is extended from approximately \SI{2.1}{Mpc} to \SI{7.3}{Mpc} in the case of $B_{\rm RMS}$ = \SI{0.1}{nG}.
This distance includes potential extragalactic source candidates like Centaurus A at about \SI{3.8}{Mpc} distance \cite{harris_rejkuba_harris_2010}.
As before, the photon horizon does not increase substantially considering a \SI{1}{nG} magnetic field.
We conclude that the unknown extragalactic magnetic field makes a reliable estimate of the UHE photon horizon challenging.\section{Summary}
\label{sec:summary}

The multi-messenger approach is becoming increasingly important in cosmic-ray physics.
In this context, CRPropa~3 was developed as a general and versatile simulation tool for the intergalactic and galactic propagation of cosmic rays.
In this paper we have presented two extensions for EM particles which are incorporated in the CRPropa 3.1 release.

The first extension concerns the photon production by cosmic-ray nuclei.
We have implemented the following new channels: photon emission following photodisintegration, elastic scattering and radiative decay as a supplement to the already existing channels of pion production, pair production and $\beta$-decay.
Using the case of UHE iron nuclei we find that photodisintegration and elastic scattering can give rise to dominant contributions in the photon emission spectrum in the PeV range for a \SI{50}{EeV} nucleus.
A comparison of the total energy budget of the individual channels yields that, while most of the energy is deposited via electron pair production, most production channels become more relevant with higher cosmic-ray energies.
Using these photon production channels in CRPropa leads to an increase in the predictive power of photon flux calculations.

The second extension enables the propagation of EM particles in CRPropa.
To this end, pair and double pair production for cosmic-ray photons as well as triplet pair production, inverse Compton scattering and synchrotron radiation for cosmic-ray electrons have been implemented.
This allows for a simulatation of EM cascades in various cosmic-ray scenarios within the CRPropa framework down to GeV energies.

As an example application we computed the photon horizon for energies above $\SI{e17}{eV}$.
Here, we simulated the EM cascade in a turbulent magnetic field, taking into account interactions with background photons, magnetic deflections and cosmological redshift.
An important new feature is the ability to take into account secondary cascade photons with small angular deflections from the initial primary photon direction. We find that for small magnetic field strengths the photon horizon can be significantly enlarged by including secondary photons.

\section*{Acknowledgments}

We thank our colleagues in the CRPropa~3 team for valuable comments and discussions.
This work is supported by the Ministry of Innovation, Science and Research of the State of North Rhine-Westphalia, and the Federal Ministry of Education and Research (BMBF).

\section*{References}

\bibliography{biblio}

\end{document}